\documentclass[conference]{IEEEtran}
\IEEEoverridecommandlockouts

\usepackage{amsmath,amssymb,amsfonts}
\usepackage{textcomp}
\usepackage{graphicx}
\usepackage[dvipsnames]{xcolor}
\usepackage{array}
\usepackage{booktabs}
\usepackage{siunitx}
\usepackage[backend=bibtex,
            language=english,
            autolang=other,
            style=ieee]{biblatex}
\usepackage[hidelinks]{hyperref}
\usepackage[capitalize]{cleveref}
\usepackage[nonumberlist]{glossaries}
\makenoidxglossaries
\newacronym{2d}{2D}{two-dimensional}
\newacronym{3d}{3D}{three-dimensional}
\newacronym{3gpp}{3GPP}{3rd generation partnership project}
\newacronym{3msk}{3MSK}{3-level minimum-shift-keying}
\newacronym{4g}{4G}{fourth-generation}
\newacronym{5g}{5G}{fifth-generation}
\newacronym{6dof}{6DoF}{six degrees of freedom}
\newacronym{6g}{6G}{sixth-generation}
\newacronym{abp}{ABP}{authentication by personalisation}
\newacronym{ac}{AC}{alternating current}
\newacronym{aci}{ACI}{adjacent channel interference}
\newacronym{aclr}{ACLR}{adjacent channel leakage ratio}
\newacronym{acpr}{ACPR}{adjacent channel power ratio}
\newacronym{adc}{ADC}{analog-to-digital converter}
\newacronym{adr}{ADR}{adaptive data rate}
\newacronym{aec}{AEC}{aluminum electrolytic capacitor}
\newacronym{aes}{AES}{Advanced Encryption Standard}
\newacronym{afe}{AFE}{analog front end}
\newacronym{ag}{AG}{array gain}
\newacronym{agc}{AGC}{automatic gain controller}
\newacronym{agps}{A-GPS}{Assisted Global Positioning System} %bestaat al :) % Aha, dan maak ik er wel Assisted gps van!
\newacronym{ai}{AI}{artificial intelligence}
\newacronym{alk}{ALK}{Alkaline}
\newacronym{am}{AM}{amplitude modulation}
\newacronym{amam}{AM/AM}{amplitude modulation to amplitude modulation}
\newacronym{amc}{AMC}{automatic modulation classification}
\newacronym{amp}{AMP}{approximate message passing}
\newacronym{ampm}{AM/PM}{amplitude modulation to phase modulation}
\newacronym{aoa}{AOA}{angle-of-arrival}
\newacronym{aod}{AOD}{angle-of-departure}
\newacronym{ap}{AP}{access point}
\newacronym{api}{API}{application program interface}
\newacronym{apt}{APT}{acoustic power transfer}
\newacronym{apu}{APU}{access point unit}
\newacronym{ar}{AR}{augmented reality}
\newacronym{arima}{ARIMA}{Auto-Regressive Integrated Moving Average}
\newacronym{arp}{ARP}{Antenna Reference Point}
\newacronym{arft}{ARFT}{acoustic-radio fusion in Techtile}
\newacronym{asic}{ASIC}{application specific integrated circuit}
\newacronym{ask}{ASK}{amplitude-shift keying}
\newacronym{at}{AT}{ATtention}
\newacronym{auv}{AUV}{autonomous underwater vehicle}
\newacronym{awg}{AWG}{arbitrary waveform generator}
\newacronym{awgn}{AWGN}{additive white Gaussian noise}
\newacronym{baw}{BAW}{bulk acoustic wave}
\newacronym{bb}{BB}{base-band}
\newacronym{bc}{BC}{backscatter communication}
\newacronym{bcjr}{BCJR}{Bahl-Cocke-Jelinek-Raviv}
\newacronym{bd}{BD}{backscatter device}
\newacronym{be}{BE}{Belgium}
\newacronym{ber}{BER}{bit error rate}
\newacronym{bf}{BF}{beamforming}
\newacronym{bga}{BGA}{ball grid array}
\newacronym{bldc}{BLDC}{brushless DC}
\newacronym{bler}{BLER}{block error rate}
\newacronym{bod}{BOD}{Brown-Out Detection}
\newacronym{bom}{BOM}{bill of materials}
\newacronym{bpsk}{BPSK}{binary phase-shift keying}
\newacronym{brp}{BRP}{beam refinement process}
\newacronym{bs}{BS}{base station}
\newacronym{bu}{BU}{booster unit}
\newacronym{bw}{BW}{bandwidth}
\newacronym{ca}{CA}{carrier emitter}
\newacronym{cad}{CAD}{channel activity detection}
\newacronym{cars}{CARS}{calibration reference signal}
\newacronym{cbm}{CBM}{condition based maintenance}
\newacronym{cc}{CC}{constant current}
\newacronym{ccdf}{CCDF}{complementary cumulative distribution function}
\newacronym{ccnn}{CCNN}{circular convolutional neural network}
\newacronym{ccs}{CCS}{correlative channel sounder}
\newacronym{cdf}{CDF}{cumulative distribution function}
\newacronym{cdma}{CDMA}{code division-multiple access}
\newacronym{cdrx}{CDRX}{connected mode DRX}
\newacronym{ce}{CE}{coverage enhancement}
\newacronym{ced}{CED}{cumulative energy density}
\newacronym{ceofdm}{CE-OFDM}{constant-envelope OFDM}
\newacronym{cf}{CF}{cell-free}
\newacronym{cfmmimo}{CF-mMIMO}{cell-free massive MIMO}
\newacronym{cfo}{CFO}{carrier frequency offset}
\newacronym{cir}{CIR}{channel impulse response}
\newacronym{cla}{CLA}{closed-loop approach}
\newacronym{clk}{CLK}{clock}
\newacronym{cmos}{CMOS}{complementary metal oxide semiconductor}
\newacronym{cnn}{CNN}{convolutional neural network}
\newacronym{co}{CO}{combinatorial optimization}
\newacronym{cordic}{CORDIC}{coordinate rotation digital computer}
\newacronym{cost}{COST}{commercial off-the-shelf}
\newacronym{cots}{COTS}{commercial off-the-shelf}
\newacronym{cp}{CP}{cyclic prefix}
\newacronym{cpe}{CPE}{common phase error}
\newacronym{cpfsk}{CPFSK}{continuous phase frequency shift keying}
\newacronym{cpm}{CPM}{continuous phase modulation}
\newacronym{cpt}{CPT}{capacitive power transfer}
\newacronym{cpu}{CPU}{central-processing unit}
\newacronym{cpw}{CPW}{coplanar waveguide}
\newacronym{cqi}{CQI}{channel quality indicator}
\newacronym{cr}{CR}{coding rate}
\newacronym{crc}{CRC}{cyclic redundancy check}
\newacronym{crlb}{CRLB}{Cram\'er-Rao lower bound}
\newacronym{crs}{CRS}{cell reference signal}
\newacronym{cs}{CS}{compressed sensing}
\newacronym{cse}{CSE}{channel state estimation}
\newacronym{csi}{CSI}{channel state information}
\newacronym{csp}{CSP}{contact service point}
\newacronym{css}{CSS}{chirp spread spectrum}
\newacronym{cu}{CU}{central unit}
\newacronym{cv}{CV}{constant voltage}
\newacronym{cw}{CW}{continuous wave}
\newacronym{d2d}{D2D}{device-to-device}
\newacronym{dab}{DAB}{Digital Audio Broadcasting}
\newacronym{dac}{DAC}{digital-to-analog converter}
\newacronym{daq}{DAQ}{data acquisition system}
\newacronym{das}{DAS}{distributed antenna systems}
\newacronym{dbpsk}{DBPSK}{Differential Binary Phase Shift Keying}
\newacronym{dc}{DC}{direct current}
\newacronym{dcc}{DCC}{dynamic cooperation clustering}
\newacronym{ddc}{DDC}{digital down conversion}
\newacronym{de}{DE}{drain efficiency}
\newacronym{dft}{DFT}{discrete Fourier transform}
\newacronym{dftsofdm}{DFT-s-OFDM}{discrete Fourier Transform spread OFDM}
\newacronym{dftsofdmfdss}{DFT-s-OFDM-FDSS}{DFT-s-OFDM with frequency-domain spectral shaping}
\newacronym{dftsofdmfdssse}{DFT-s-OFDM-FDSS-SE}{DFT-s-OFDM with FDSS and spectral extension}
\newacronym{dl}{DL}{downlink}
\newacronym{dlc}{DLC}{Distributed Laser Charging}
\newacronym{dli}{DLI}{direct link interference}
\newacronym{dlt}{DLT}{Distributed Ledger Technology}
\newacronym{dm}{DM}{diffuse multipath}
\newacronym{dma}{DMA}{Direct Memory Access}
\newacronym{dmac}{DMAC}{Direct Memory Access Controller}
\newacronym{dmc}{DMC}{diffuse multipath component}
\newacronym{dmimo}{D-MIMO}{distributed MIMO}
\newacronym{dnn}{DNN}{deep neural network}
\newacronym{doa}{DOA}{direction-of-arrival}
\newacronym{dp}{DP}{differencial privacy}
\newacronym{dpd}{DPD}{digital pre-distortion}
\newacronym{dpdk}{DPDK}{Data Plane Development Kit}
\newacronym{dram}{DRAM}{dynamic random-access memory}
\newacronym{drcs}{$\Delta$RCS}{differential-radar cross section}
\newacronym{drx}{DRX}{Discontinuous Reception Mode}
\newacronym{dsb}{DSB}{double-sideband}
\newacronym{dsl}{DSL}{digital subscriber line}
\newacronym{dsp}{DSP}{digital signal processing}
\newacronym{dss}{DSS}{Dataset Storage Standard}
\newacronym{duc}{DUC}{digital up-converter}
\newacronym{dvb}{DVB}{Digital Video Broadcasting}
\newacronym{e2e}{E2E}{end-to-end}
\newacronym{easa}{EASA}{European Union Aviation Safety Agency}
\newacronym{ebg}{EBG}{electromagnetic bandgap}
\newacronym{ebw}{EBW}{excess bandwidth}
\newacronym{ec}{EC}{European Commission}
\newacronym{ecc}{ECC}{elliptic curve cryptography}
\newacronym{ecdf}{eCDF}{empirical cumulative distribution function}
\newacronym{ecdlp}{ECDLP}{elliptic curve discrete logarithm problem}
\newacronym{ecsp}{ECSP}{edge computing service point}
\newacronym{edlc}{EDLC}{electrostatic double-layer capacitors}
\newacronym{edrx}{eDRX}{Extended Discontinuous Reception Mode}
\newacronym{ee}{EE}{energy efficiency}
\newacronym{egprs}{EGPRS}{Enhanced Data Rates for GSM Evolution}
\newacronym{eh}{EH}{energy harvesting}
\newacronym{eirp}{EIRP}{equivalent isotropically radiated power}
\newacronym{em}{EM}{electromagnetic}
\newacronym{embb}{eMBB}{enhanced Mobile Broadband}
\newacronym{en}{EN}{energy neutral}
\newacronym{end}{END}{energy-neutral device}
\newacronym{enob}{ENOB}{effective number of bits}
\newacronym{eol}{EoL}{end of life}
\newacronym{ep}{EP}{energy profiler}
\newacronym{epd}{EPD}{electronic paper display}
\newacronym{epu}{EPU}{edge processing unit}
\newacronym{er}{ER}{Energy Receiver}
\newacronym{erp}{ERP}{effective radiation power}
\newacronym[plural=ESCs,firstplural=electronic speed controllers (ESCs)]{esc}{ESC}{electronic speed control}
\newacronym{esd}{ESD}{electrostatic discharge}
\newacronym{esl}{ESL}{electronic shelf label}
\newacronym{esr}{ESR}{equivalent series resistance}
\newacronym{et}{ET}{Energy Transmitter}
\newacronym{etsi}{ETSI}{European Telecommunications Standards Institute}
\newacronym{evd}{EVD}{eigenvalue decomposition}
\newacronym{evm}{EVM}{Error Vector Magnitude}
\newacronym{ewlb}{eWLB}{Embedded Wafer Level Ball Grid Array}
\newacronym{fa}{FA}{federation anchor}
\newacronym{fair}{FAIR}{Findability, Accessibility, Interoperability, and Reuse of digital assets}
\newacronym{fcf}{FCF}{frequency correlation function}
\newacronym{fd}{FD}{front-haul distance}
\newacronym{fdd}{FDD}{frequency-division duplexing}
\newacronym{fde}{FDE}{frequency domain equalizer}
\newacronym{fdm}{FDM}{frequency-division multiplexing}
\newacronym{fdma}{FDMA}{frequency division multiple access}
\newacronym{fdss}{FDSS}{frequency-domain spectral shaping}
\newacronym{fem}{FEM}{finite element analysis}
\newacronym{fembb}{feMBB}{further enhanced mobile broadband}
\newacronym{fft}{FFT}{fast Fourier transform}
\newacronym{fh}{FH}{fronthaul}
\newacronym{fhss}{FHSS}{frequency hopping spread spectrum}
\newacronym{fifo}{FIFO}{First In, First Out}
\newacronym{fim}{FIM}{Fisher information matrix}
\newacronym{fits}{FITS}{Flexible Image Transport System}
\newacronym{fl}{FL}{federated learning}
\newacronym{fom}{FoM}{figure of merit}
\newacronym{fov}{FOV}{field of view}
\newacronym{fpga}{FPGA}{field-programmable gate array}
\newacronym{fr1}{FR1}{frequency range 1}
\newacronym{fr2}{FR2}{frequency range 2}
\newacronym{fram}{FRAM}{Ferroelectric Random Access Memory}
\newacronym{fsk}{FSK}{frequency shift keying}
\newacronym{fspl}{FSPL}{free space path loss}
\newacronym{fss}{FSS}{frequency selective surface}
\newacronym{gb}{GB}{grant-based}
\newacronym{gdpr}{GDPR}{general data protection regulation}
\newacronym{gf}{GF}{grant-free}
\newacronym{gmsk}{GMSK}{Gaussian minimum-shift keying}
\newacronym{gnb}{gNB}{Next Generation Node B}
\newacronym{gni}{GNI}{gross national income}
\newacronym{gnn}{GNN}{graph neural network}
\newacronym{gnss}{GNSS}{global navigation satellite system}
\newacronym{gpclk}{GPCLK}{general purpose clock}
\newacronym{gpio}{GPIO}{General-Purpose Input/Output}
\newacronym{gpl}{GPL}{GNU General Public License}
\newacronym{gprs}{GPRS}{General Packet Radio Services}
\newacronym{gps}{GPS}{Global Positioning System}
\newacronym{gpu}{GPU}{graphical processing unit}
\newacronym{grc}{GRC}{GNU Radio Companion}
\newacronym{gscm}{GSCM}{geometry‐based stochastic model}
\newacronym{gsm}{GSM}{Global System for Mobile Communications}  %
\newacronym{gspm}{GSpM}{Generalized spatial modulation}
\newacronym{gwp}{GWP}{Global Warming Potential}
\newacronym{harq}{HARQ}{hybrid automatic repeat request}
\newacronym{hat}{HAT}{hardware attached on top}
\newacronym{hcs}{HCS}{human-centric services}
\newacronym{hdf5}{HDF5}{Hierarchical Data Format version 5}
\newacronym{hfss}{HFSS}{High Frequency Simulator Software}
\newacronym{hmd}{HMD}{head-mounted display}
\newacronym{hpbm}{HPBM}{half power beam width}
\newacronym{HyMPRo}{HyMPRo}{Hybrid Multi-Path Routing algorithm}
\newacronym{i.i.d.}{i.i.d.}{independent and identically distributed}
\newacronym{i2c}{I2C}{Inter-Integrated Circuit}
\newacronym{iaq}{IAQ}{Indoor Air Quality}
\newacronym{ib}{IB}{in-band}
\newacronym{ibbc}{IBBC}{inter-band beam configuration}
\newacronym{ibo}{IBO}{input back-off}
\newacronym{ic}{IC}{integrated circuit}
\newacronym{iccs}{ICCS}{Ilmsens correlative channel sounder}
\newacronym{ici}{ICI}{intercarrier interference}
\newacronym{icnirp}{ICNIRP}{International Commission on Non-Ionizing Radiation Protection}
\newacronym{id}{ID}{information decoding}
\newacronym{idft}{IDFT}{inverse discrete Fourier transform}
\newacronym{idxm}{IDXM}{index modulation}
\newacronym{if}{IF}{intermediate-frequency}
\newacronym{ifft}{IFFT}{inverse fast-Fourier-transform}
\newacronym{iid}{i.i.d.}{independently and identically distributed}
\newacronym{iis}{IIS}{integrated information system}
\newacronym{im}{IM}{intermodulation}
\newacronym{imd}{IMD}{intermodulation distortion}
\newacronym{imu}{IMU}{inertial measurement unit}
\newacronym{inh}{InH}{indoor hotspot office}
\newacronym{io}{IO}{input/output}
\newacronym{ioe}{IoE}{Internet of Everything}
\newacronym{iot}{IoT}{Internet of Things}
\newacronym{ipt}{IPT}{inductive power transfer}
\newacronym{ipy}{IPY}{Interventions per Year}
\newacronym{iq}{IQ}{in-phase and quadrature}
\newacronym{iqi}{IQI}{IQ imbalance}
\newacronym{ir}{IR}{infrared}
\newacronym{isi}{ISI}{intersymbol interference}
\newacronym{ism}{ISM}{industrial, scientific and medical}
\newacronym{isp}{ISP}{internet service provider}
\newacronym{itu}{ITU}{International Telecommunication Union}
\newacronym{jesd}{JESD}{Joint Electron Devices Engineering Council}
\newacronym{jfet}{JFET}{junction field effect transistor}
\newacronym{kpi}{KPI}{key performance indicator}
\newacronym{ktofdm}{KT-DFT-s-OFDM}{known-tail-DFT-s-OFDM}
\newacronym{kvi}{KVI}{key value indicator}
\newacronym{larva}{LARVA}{LARge Virtual Array}
\newacronym{lca}{LCA}{life cycle assessment}
\newacronym{lco}{LCO}{lithium cobalt oxide}
\newacronym{ldo}{LDO}{Low-dropout voltage regulator}
\newacronym{ldpc}{LDPC}{low-density parity-check}
\newacronym{led}{LED}{Light Emitting Diode}
\newacronym{less}{LESS}{Low Energy Scheduler Solution}
\newacronym{lfp}{LFP}{lithium iron phosphate}
\newacronym{lib}{LIB}{Lithium-Ion Battery}
\newacronym{lic}{LIC}{lithium-ion capacitor}
\newacronym{lid}{LID}{Lithium Iron Disulfide}
\newacronym{lidar}{LiDAR}{light detection and ranging}
\newacronym{liion}{Li-ion}{lithium-ion}
\newacronym{lipo}{LiPo}{lithium polymer}
\newacronym{lis}{LIS}{large intelligent surface}
\newacronym{llh}{LLH}{log-likelihood}
\newacronym{lls}{LLS}{link-level simulation}
\newacronym{lmd}{LMD}{Lithium Manganese Dioxide}
\newacronym{lmmse}{LMMSE}{least minimum mean square error}
\newacronym{lmo}{LMO}{lithium ion manganese oxide}
\newacronym{lna}{LNA}{low-noise amplifier}
\newacronym{lo}{LO}{local oscillator}
\newacronym{lora}{LoRa}{long range}
\newacronym{lorawan}{LoRaWAN}{long-range wide-area network}
\newacronym{los}{LoS}{line-of-sight}
\newacronym{lp}{LP}{linear programming}
\newacronym{lpf}{LPF}{low-pass filter}
\newacronym{lpt}{LPT}{laser power transfer}
\newacronym{lpwa}{LPWA}{Low Power Wide Area}
\newacronym{lpwan}{LPWAN}{low-power wide-area network}
\newacronym{lpwans}{LPWANs}{Low-Power Wide-Area Networks}
\newacronym{lqi}{LQI}{link quality indicator}
\newacronym{lrelu}{LReLU}{leaky rectified linear unit}
\newacronym{lrt}{LRT}{likelihood-ratio test}
\newacronym{ls}{LS}{least squares}
\newacronym{lsa}{LSA}{large synthetic array}
\newacronym{lsf}{LSF}{large-scale fading}
\newacronym{lsfc}{LSFC}{large-scale fading component}
\newacronym{lstm}{LSTM}{Long Short-Term Memory}
\newacronym{ltc}{LTC}{lithium thionyl chloride}
\newacronym{lte}{LTE}{Long Term Evolution}
\newacronym{lti}{LTI}{linear time-invariant}
\newacronym{lto}{LTO}{lithium titanate}
\newacronym{lusta}{LUSTA 5G }{Logistique mUltimodale Sécuritaire Téléopérée \& Autonome 5G}
\newacronym{m2m}{M2M}{machine to machine}
\newacronym{mac}{MAC}{Medium Access Control}
\newacronym{mate}{MATE}{millimeter-wave MIMO testbed}
\newacronym{mc}{MC}{Monte Carlo}
\newacronym{mcl}{MCL}{Maximum Coupling Loss}
\newacronym{mcs}{MCS}{modulation and coding scheme}
\newacronym{mcu}{MCU}{microcontroller unit}
\newacronym{mec}{MEC}{multi-access edge computing}
\newacronym{mems}{MEMS}{micro-electromechanical systems}
\newacronym{mf}{MF}{matched filter}
\newacronym{mimo}{MIMO}{multiple-input multiple-output}
\newacronym{miso}{MISO}{multiple-input single-output}
\newacronym{ml}{ML}{machine learning}
\newacronym{mlp}{MLP}{multilayer perceptron}
\newacronym{mmic}{MMIC}{monolithic microwave integrated circuit}
\newacronym{mmimo}{mMIMO}{massive MIMO}
\newacronym{mmse}{MMSE}{minimum mean square error}
\newacronym{mmtc}{mMTC}{massive machine-typed communication}
\newacronym{mmwave}{mmWave}{millimeter wave}
\newacronym{mn}{MN}{matching network}
\newacronym{mosfet}{MOSFET}{metal-oxide semiconductor field effect transistor}
\newacronym{mpc}{MPC}{multipath component}
\newacronym{mppt}{MPPT}{maximum power point tracking}
\newacronym{mr}{MR}{maximum ratio}
\newacronym{mrc}{MRC}{maximum ratio combining}
\newacronym{mrc_em}{MRC}{maximum ratio combining}
\newacronym{mrc_EM}{MRC}{Magnetic Resonance Coupling}
\newacronym{mrt}{MRT}{maximum ratio transmission}
\newacronym{mse}{MSE}{mean square error}
\newacronym{msk}{MSK}{Minimum-Shift Keying}
\newacronym{mtc}{MTC}{Machine-Type Communication}
\newacronym{multi-rat}{Multi-RAT}{multiple radio access technology}
\newacronym{multirat}{Multi-RAT}{Multiple Radio Access Technology}
\newacronym{music}{MUSIC}{MUltiple SIgnal Classification}
\newacronym{navauwall}{NAVAUWALL}{AUtomated NAVigation in WALLonia}
\newacronym{nb}{NB}{narrowband}
\newacronym{nbiot}{NB-IoT}{narrowband IoT}
\newacronym{nca}{NCA}{nickel cobalt aluminum}
\newacronym{netcdf}{NetCDF}{Network Common Data Form}
\newacronym{nf}{NF}{noise figure}
\newacronym{nfc}{NFC}{near-field communication}
\newacronym{nfv}{NFV}{network function virtualization}
\newacronym{ngmn}{NGMN}{Next Generation Mobile Networks }
\newacronym{ni}{NI}{National Instruments}
\newacronym{nicd}{NiCd}{nikkel cadmium}
\newacronym{nimh}{NiMH}{nikkel metal hydride}
\newacronym{nlos}{NLoS}{non-line-of-sight}
\newacronym{nmc}{NMC}{nickel manganese cobalt}
\newacronym{nmos}{nMOS}{n-channel metal-oxide semiconductor}
\newacronym{nn}{NN}{neural network}
\newacronym{nnls}{NNLS}{non-negative least squares}
\newacronym{noma}{NOMA}{non-orthogonal multiple access}
\newacronym{np}{NP}{Neyman-Pearson}
\newacronym{npbch}{NPBCH}{Narrowband Physical Broadcast Channel}
\newacronym{npss}{NPSS}{Narrow Band Primay Synchronization Signal}
\newacronym{nr}{NR}{New Radio}
\newacronym{nrs}{NRS}{Narrow Band Reference Signal}
\newacronym{nsss}{NSSS}{Narrowband Secondary Synchronization Signal}
\newacronym{ntp}{NTP}{network time protocol}
\newacronym{oai}{OAI}{OpenAirInterface} % no spelling mistake, it is spelled all glued to eachother...
\newacronym{obw}{OBW}{occupied bandwidth}
\newacronym{ofdm}{OFDM}{orthogonal frequency-division multiplexing}
\newacronym{ofdma}{OFDMA}{orthogonal frequency-division multiple access}
\newacronym{ofdmim}{OFDM-IM}{OFDM with index modulation}
\newacronym{olos}{OLoS}{obstructed-line-of-sight}
\newacronym{oma}{OMA}{orthogonal multiple access}
\newacronym{oob}{OOB}{out-of-band}
\newacronym{ook}{OOK}{on-off keying}
\newacronym{opbo}{OPBO}{output power backoff}
\newacronym{oran}{O-RAN}{open radio-access network}
\newacronym{os}{OS}{operating system}
\newacronym{ota}{OTA}{over-the-air}
\newacronym{otaa}{OTAA}{over-the-air authentication}
\newacronym{p1}{P1}{Phase 1}
\newacronym{p2}{P2}{Phase 2}
\newacronym{p2p}{P2P}{point-to-point}
\newacronym{pa}{PA}{power amplifier}
\newacronym{pae}{PAE}{power-added efficiency}
\newacronym{pam}{PAM}{pulse amplitude modulation}
\newacronym{pana}{PanA}{Panel A}
\newacronym{panb}{PanB}{Panel B}
\newacronym{papr}{PAPR}{peak-to-average power ratio}
\newacronym{pc}{PC}{pilot count}
\newacronym{pcb}{PCB}{printed circuit board}
\newacronym{pcg}{PCG}{power consumption gain}
\newacronym{pcie}{PCIe}{Peripheral Component Interconnect Express}
\newacronym{pcsi}{PCSI}{perfect channel state information}
\newacronym{pd}{PD}{powered device}
\newacronym{pdcch}{PDCCH}{physical downlink control channel}
\newacronym{pdf}{PDF}{probability density function}
\newacronym{pdp}{PDP}{power delay profile}
\newacronym{pdsch}{PDSCH}{physical downlink shared channel}
\newacronym{pe}{PE}{processing element}
\newacronym{peb}{PEB}{positioning error bound}
\newacronym{per}{PER}{packet error rate}
\newacronym{pet}{PET}{privacy enhancing technology}
\newacronym{pg}{PG}{path gain}
\newacronym{pgd}{PGD}{proximal gradient descent}
\newacronym{phy}{PHY}{physical}
\newacronym{pki}{PKI}{public key infrastucture}
\newacronym{pl}{PL}{path loss}
\newacronym{pla}{PLA}{physically large array}
\newacronym{pll}{PLL}{phase-locked loop}
\newacronym[plural=PMs,firstplural=person months (PMs)]{pm}{PM}{person month}
\newacronym{pmf}{PMF}{polymer microwave fiber}
\newacronym{pmu}{PMU}{power management unit}
\newacronym{pn}{PN}{pseudo-noise}
\newacronym{po}{PO}{phase offset}
\newacronym{poc}{PoC}{proof of concept}
\newacronym{poe}{PoE}{power-over-Ethernet}
\newacronym{pps}{1PPS}{pulse per second}
\newacronym{pr}{PR}{phase reversal}
\newacronym{prb}{PRB}{Physical Resource Block}
\newacronym{prbs}{PRBs}{Physical Resource Blocks}
\newacronym{prs}{PRS}{Peripheral Reflex System}
\newacronym{ps}{PS}{Processing System}
\newacronym{psd}{PSD}{power spectral density}
\newacronym{pse}{PSE}{power sourcing equipment}
\newacronym{psk}{PSK}{phase shift keying}
\newacronym{psm}{PSM}{power saving mode}
\newacronym{pss}{PSS}{primary synchronisation signal}
\newacronym{ptp}{PTP}{precision-time protocol}
\newacronym{ptrs}{PTRS}{Phase-Tracking Reference Signals}
\newacronym{ptw}{PTW}{paging time window}
\newacronym{pv}{PV}{photovoltaic}
\newacronym{pw}{PW}{planar wavefront}
\newacronym{pwm}{PWM}{pulse width modulation}
\newacronym{qam}{QAM}{quadrature amplitude modulation}
\newacronym{qos}{QoS}{quality-of-service}
\newacronym{qpsk}{QPSK}{quadrature phase-shift keying}
\newacronym{qrrls}{QR-RLS}{QR decomposition based recursive least squares}
\newacronym{quadriga}{QuaDRiGa}{QUAsi Deterministic RadIo channel GenerAtor}
\newacronym{ra}{RA}{Random Access}
\newacronym{ram}{RAM}{random-access memory}
\newacronym{ran}{RAN}{radio access network}
\newacronym{rar}{RAR}{Random Access Response}
\newacronym{rat}{RAT}{radio access technology}
\newacronym{rb}{Rb}{Rubidium}
\newacronym{rbs}{RBS}{radio base station}
\newacronym{rbw}{RBW}{resolution bandwidth}
\newacronym{rc}{RC}{raised-cosine}
\newacronym{rcs}{RCS}{radar cross section}
\newacronym{rdl}{RDL}{redistribution layer}
\newacronym{re}{RE}{radio element}
\newacronym{relu}{ReLU}{rectified linear unit}
\newacronym{rf}{RF}{radio frequency}
\newacronym{rfeh}{RFEH}{radio frequency energy harvesting}
\newacronym{rfic}{RFIC}{radio-frequency integrated circuit}
\newacronym{rfid}{RFID}{radio frequency identification}
\newacronym{rfpt}{RFPT}{radio frequency power transfer}
\newacronym{rfsoc}{RFSoC}{Radio Frequency System-on-Chip}
\newacronym{rir}{RIR}{room impulse response}
\newacronym{ris}{RIS}{reflective intelligent surface}%reconfigurable?
\newacronym{rllmtc}{RLLMTC}{reliable low latency machine type communication}
\newacronym{rls}{RLS}{recursive least squares}
\newacronym{rms}{RMS}{root-mean-square}
\newacronym{rmse}{RMSE}{root-mean-square error}
\newacronym{rmt}{RMT}{random matrix theory}
\newacronym{rof}{RoF}{radio-over-fiber}
\newacronym{ros}{ROS}{robot operating system}
\newacronym{rpi}{RPi}{Raspberry Pi}
\newacronym{rrc}{RRC}{Radio Resource Connection}
\newacronym{rreq}{RREQ}{route request packet}
\newacronym{rsrp}{RSRP}{Reference Signals Received Power}
\newacronym{rsrq}{RSRQ}{Reference Signal Received Quality}
\newacronym{rss}{RSS}{received signal strength}
\newacronym{rssi}{RSSI}{received signal strength indicator}
\newacronym{rtc}{RTC}{real time clock}
\newacronym{rtf}{RTF}{reader talks first}
\newacronym{rtk}{RTK}{real time kinematics}
\newacronym{rts}{RTS}{ray tracing simulator}
\newacronym{ru}{RU}{Radio Unit}
\newacronym{rv}{RV}{random variable}
\newacronym{rw}{RW}{RadioWeaves}
\newacronym{rx}{RX}{receiver}
\newacronym{rzf}{RZF}{regularized zero forcing}
\newacronym{s-parameter}{S-parameter}{scattering parameter}
\newacronym{sa}{SA}{synchronization anchor}
\newacronym{sa5}{SA}{Stand Alone}
\newacronym{sar}{SAR}{specific absorption rate}
\newacronym{sbl}{SBL}{sparse Bayesian learning}
\newacronym{sc}{SC}{single carrier}
\newacronym{scfde}{SC-FDE}{single-carrier modulation with frequency-domain-equalization}
\newacronym{scfdma}{SCFDMA}{single-carrier frequency division multiple access}
\newacronym{scs}{SCS}{sub-carrier spacing}
\newacronym{sdg}{SDG}{Sustainable Development Goal}
\newacronym{sdm}{SDM}{sigma-delta modulator}
\newacronym{sdma}{SDMA}{spatial-division multiple access}
\newacronym{sdn}{SDN}{software-defined network}
\newacronym{sdof}{SDoF}{sigma-delta over fiber}
\newacronym{sdr}{SDR}{software-defined radio}
\newacronym{se}{SE}{spectral efficiency}
\newacronym{sei}{SEI}{specific emitter identification}
\newacronym{ser}{SER}{symbol-error rate}
\newacronym{sf}{SF}{spreading factor}
\newacronym{sfn}{SFN}{single frequency network}
\newacronym{sfp}{SFP}{small form-factor pluggable}
\newacronym{sha}{SHA}{Secure Hash Algorithm}
\newacronym{SigMF}{SigMF}{Signal Metadata Format}
\newacronym{simo}{SIMO}{single-input multiple-output}
\newacronym{sinr}{SINR}{signal-to-interference-plus-noise ratio}
\newacronym{siso}{SISO}{single-input single-output}
\newacronym{slam}{SLAM}{simultaneous localization and mapping}
\newacronym{slc}{SLC}{spatial leakage suppression}
\newacronym{slerp}{SLERP}{spherical linear interpolation}
\newacronym{sma}{SMA}{SubMiniature version A}
\newacronym{smc}{SMC}{specular multipath component}
\newacronym{smps}{SMPS}{switched mode power supply}
\newacronym{sndr}{SNDR}{signal-to-noise-and-distortion ratio}
\newacronym{snidr}{SNIDR}{signal-to-noise-and-interference-and-distortion ratio}
\newacronym{snir}{SNIR}{signal-to-interference-plus-noise ratio}
\newacronym{snr}{SNR}{signal-to-noise ratio}
\newacronym{soc}{SoC}{state of charge}
\newacronym{SoC}{SOC}{System on Chip}
\newacronym{sp}{SP}{service point}
\newacronym{spdt}{SPDT}{single pole double throw}
\newacronym{spi}{SPI}{Serial Peripheral Interface}
\newacronym{spst}{SPST}{single pole single throw}
\newacronym{sram}{SRAM}{static random-access memory}
\newacronym{srd}{SRD}{short-range device}
\newacronym{srls}{SRLS}{standard recursive least squares}
\newacronym{srs}{SRS}{Sounding Reference Signal}
\newacronym{ssb}{SSB}{synchronisation signal block}
\newacronym{ssd}{SSD}{solid state drive}
\newacronym{ssq}{SSQ}{simulator sickness questionnaire}
\newacronym{steam}{STEAM}{science, technology, engineering, the arts, and mathematics}
\newacronym{svd}{SVD}{singular value decomposition}
\newacronym{sw}{SW}{spherical wavefront}
\newacronym{swipt}{SWIPT}{simultaneous wireless information and power transfer}
\newacronym{synce}{SyncE}{Synchronous Ethernet}
\newacronym{ta}{TA}{timing advance}
\newacronym{tau}{TAU}{tracking area update}
\newacronym{tcer}{TCER}{transported to consumed energy ratio}
\newacronym{tcp}{TCP}{Transmission Control Protocol}
\newacronym{tcxo}{TCXO}{temperature compensated crystal oscillator}
\newacronym{tdce}{TD-CE}{time-domain compression and expansion}
\newacronym{tdd}{TDD}{time division duplexing}
\newacronym{tdma}{TDMA}{time division-multiple access}
\newacronym{tdoa}{TDOA}{time-difference-of-arrival}
\newacronym{thz}{THz}{Terahertz}
\newacronym{to}{TO}{timing offset}
\newacronym{toa}{TOA}{time-of-arrival}
\newacronym{tof}{ToF}{time-of-flight}
\newacronym{tosm}{TOSM}{through-open-short-match}
\newacronym{tpms}{TPMS}{Tire-Pressure Monitoring System}
\newacronym{trl}{TRL}{technology readyness level}
\newacronym{trp}{TRP}{Transmission Reception Point}
\newacronym{tsn}{TSN}{time-sensitive networking}
\newacronym{ttf}{TTF}{tag talks first}
\newacronym{ttff}{TTFF}{Time To First Fix}
\newacronym{tti}{TTI}{transmission time interval}
\newacronym{ttm}{TTM}{time to market}
\newacronym{ttn}{TTN}{The Things Network}
\newacronym{tx}{TX}{transmitter}
\newacronym{uart}{UART}{Universal Asynchronous Receiver/Transmitter}
\newacronym{uav}{UAV}{unmanned aerial vehicle}
\newacronym{uc}{UC}{use case}
\newacronym{ucie}{UCIe}{Universal Chiplet Interconnect Express}
\newacronym{udp}{UDP}{User Datagram Protocol}
\newacronym{ue}{UE}{user equipment}
\newacronym{ugv}{UGV}{unmanned ground vehicle}
\newacronym{uhd}{UHD}{USRP hardware driver}
\newacronym{uhf}{UHF}{ultra-high frequency}
\newacronym{ul}{UL}{uplink}
\newacronym{ula}{ULA}{uniform linear array}
\newacronym{uMIMO}{$\mu$-MIMO}{ultra-massive Multiple-Input Multiple-Output}
\newacronym{ummtc}{umMTC}{ultra massive machine type communication}
\newacronym{un}{UN}{United Nations}
\newacronym{unow}{uNOW}{unified non-orthogonal waveform}
\newacronym{upa}{UPA}{uniform planar array}
\newacronym{ura}{URA}{uniform rectangular array}
\newacronym{urllc}{URLLC}{ultra-reliable low-latency communications}
\newacronym{usrp}{USRP}{universal software radio peripheral}
\newacronym{uv}{UV}{unmanned vehicle}
\newacronym{uw}{UW}{unique word}
\newacronym{uwb}{UWB}{ultrawideband}
\newacronym{uwofdm}{UW-DFT-s-OFDM}{unique word DFT-s-OFDM}
\newacronym{v2v}{V2V}{vehicle-to-vehicle}
\newacronym{vco}{VCO}{voltage-controlled oscillator}
\newacronym{vep}{VEP}{virtual edge platform}
\newacronym{vlc}{VLC}{visible light communication}
\newacronym{vlp}{VLP}{visible light positioning}
\newacronym{vna}{VNA}{vector network analyzer}
\newacronym{voc}{VOC}{Voltatile Organic Compound}
\newacronym{votable}{VOTable}{Virtual Observatory Table}
\newacronym{vr}{VR}{virtual reality}
\newacronym{vuca}{VUCA}{volatile, uncertain, complex and ambiguous}
\newacronym{wb}{WB}{wideband}
\newacronym{wban}{WBAN}{wireless body area network}
\newacronym{wd}{WD}{Wireless distance}
\newacronym{wimax}{WiMAX}{Worldwide Interoperability for Microwave Access}
\newacronym{wlan}{WLAN}{wireless LAN}
\newacronym[plural=WPs,firstplural=work packages (WPs)]{wp}{WP}{work package}
\newacronym{wpt}{WPT}{wireless power transfer}
\newacronym{wr}{WR}{White Rabbit}
\newacronym{wrsn}{WRSN}{wireless rechargeable sensor network}
\newacronym{wsn}{WSN}{wireless sensor network}
\newacronym{xets}{XETS}{cross exponentially tapered slot}
\newacronym{xlmimo}{XL-MIMO}{extremely large-scale MIMO}
\newacronym{xr}{XR}{extended reality}
\newacronym{z3ro}{Z3RO}{zero third-order distortion}
\newacronym{zf}{ZF}{zero-forcing}
\newacronym{zmcscg}{ZMCSCG}{zero mean circularly symmetric complex Gaussian}
\newacronym{zmq}{ZMQ}{ZeroMQ}
\newacronym{ztofdm}{ZT-DFT-s-OFDM}{zero-tail DFT-s-OFDM}
\newacronym{llr}{LLR}{log-likelihood ratio}
\usepackage[font=small]{caption}
\usepackage{subcaption}
\AtBeginBibliography{\footnotesize}

\def\BibTeX{{\rm B\kern-.05em{\sc i\kern-.025em b}\kern-.08em
    T\kern-.1667em\lower.7ex\hbox{E}\kern-.125emX}}

\makeatletter
\PassOptionsToPackage{dvipsnames,svgnames,table}{xcolor}%
\makeatother

\usepackage{listofitems}
\usepackage{pgffor}
\usepackage{xparse}
\usepackage{ifthen}
\usepackage{tikz}
\usepackage{xspace}
\usetikzlibrary{external}

\makeatletter
\def\myColorList{LimeGreen,BrickRed,Fuchsia,Bittersweet,YellowOrange, YellowGreen, WildStrawberry, Fuchsia, RedViolet, Periwinkle, PineGreen, OrangeRed, RawSienna, Turquoise, Aquamarine, BlueViolet, BurntOrange, DarkOrchid, TealBlue, Violet, RoyalPurple,%
LimeGreen,BrickRed,Fuchsia,Bittersweet,YellowOrange, YellowGreen, WildStrawberry, Fuchsia, RedViolet, Periwinkle, PineGreen, OrangeRed, RawSienna, Turquoise, Aquamarine, BlueViolet, BurntOrange, DarkOrchid, TealBlue, Violet, RoyalPurple}
\readlist\ColorList\myColorList

\definecolor{defaultColor}{RGB}{112,45,128}
\NewDocumentCommand{\roundLabel}{O{defaultColor} m}{%
\tikzexternaldisable
    \tikz[baseline=(char.base)]{
        \node[rectangle, rounded corners=0.65mm, inner sep=0.65mm, fill=#1, draw=#1, text=white, font=\itshape](char) {#2};
    }%
\tikzexternalenable
}

\NewDocumentCommand{\defineauthors}{ O{true} m }{%
    \ifthenelse{\equal{#1}{true}}{%
        Personal commands for comments are:
        \begin{itemize}
        \foreach \x [count=\xi from 1] in {#2} {   
            \item \textcolor{\ColorList[\xi]}{\textbackslash\unexpanded\expandafter{\x}\{\}}
            }
        \end{itemize}
        \foreach \x [count=\xi from 1] in {#2} { 
            \expandafter\xdef\csname\x\endcsname####1{\noexpand\textcolor{\ColorList[\xi]}{\roundLabel[\ColorList[\xi]]{\unexpanded\expandafter{\x}} ####1}}%
        }
    }{%
        \foreach \x [count=\xi from 1] in {#2} { 
            \expandafter\xdef\csname\x\endcsname####1{\noexpand\textcolor{\ColorList[\xi]}{\roundLabel[\ColorList[\xi]]{\unexpanded\expandafter{\x}} ####1}}%
        }
    }
    \foreach \x [count=\xi from 1] in {#2} {
        \expandafter\xdef\csname at\x\endcsname####1{\noexpand\textcolor{\ColorList[\xi]}{@\x}}%
    }
}

\NewDocumentCommand{\removeauthors}{ m }{%
    \foreach \x in {#1} { 
        \expandafter\edef\csname\x\endcsname####1{}%
    }
    \foreach \x in {#1} {
        \expandafter\edef\csname at\x\endcsname####1{}%
    }
}

\makeatother

\usepackage{orcidlink}

\defineauthors[false]{gilles, jarne, daan, bert, gustav}

\usepackage{array}
\newcolumntype{H}{>{\setbox0=\hbox\bgroup}c<{\egroup}@{}}

\usepackage{fontawesome}
    
\begin{document}

\title{ARFT: A Synchronized Multimodal RF-Acoustic Dataset for Positioning in Distributed Environments}

%ARFT (Acoustic-Radio Fusion at Techtile)

\author{
%%%% AUTHORS %%%%
\IEEEauthorblockN{%
Daan Delabie\,\orcidlink{0000-0002-0089-3633}\IEEEauthorrefmark{1},
Jarne Van Mulders\,\orcidlink{0000-0003-4103-3290}\IEEEauthorrefmark{1},
Bert Pyck\,\IEEEauthorrefmark{1},
Gustav Nilsson Gisleskog\,\orcidlink{0009-0006-3230-8319}\IEEEauthorrefmark{2},
Gilles Callebaut\,\orcidlink{0000-0003-2413-986X}\IEEEauthorrefmark{1},
}
%%%% AFFILIATIONS %%%%% ...
\IEEEauthorblockA{\IEEEauthorrefmark{1}% 2nd affiliations
KU Leuven, Department of Electrical Engineering, KU Leuven Ghent, Belgium}
\IEEEauthorblockA{\IEEEauthorrefmark{2}% 2nd affiliations
Lund Univerisity, Centre for Mathematical Sciences, CVML, Lund, Sweden} %\\
}

\maketitle

\begin{abstract}
This paper documents the \gls{arft} dataset, a synchronized measurement campaign for distributed wireless sensing and positioning in the Techtile testbed. Ultrasonic and \gls{rf} signals are simultaneously transmitted and captured at multiple positions in a 2D spatial grid inside the Techtile testbed.
Each acquisition cycle corresponds to one rover stop, one position sample, one acoustic recording and one \gls{rf} snapshot. The \gls{rf} modality is recorded as per-host pilot measurements and released as \gls{csi} tensors for 42 antennas mounted at the ceiling of the room. The acoustic chirp is recorded with 91 synchronized microphones and transmitted by a synchronized quasi omni-directional speaker. The campaign spans 5011 spatial position samples spanning  a 5.57~m by 2.89~m area with complete RF and acoustic data. We detail how the measurements are recorded, quantify per-experiment coverage, describe the RF and acoustic data contents, and explain how the aligned modalities support RF-only, acoustic-only, and joint positioning workflows. Furthermore, a static acoustic positioning pipeline that performs anchor selection, pulse-compression ranging, and \gls{ls} localization is elaborated. 
\end{abstract}


% \tableofcontents

\section{Introduction}
\glsresetall

Large-scale experimental datasets have become increasingly important for wireless communication, sensing and positioning research because they decouple algorithm development from scarce hardware testbed time and expose downstream users to realistic synchronization errors, coverage gaps, and multimodal alignment constraints. Recent multimodal releases such as DeepSense~6G~\cite{alkhateeb2023deepsense} demonstrate how combining wireless measurements with complementary sensing modalities can accelerate communication-aware sensing research. Public \gls{rf} localization datasets have established important baselines for \gls{csi} fingerprinting, Wi-Fi positioning and distributed MIMO localization~\cite{kazemi2022csifingerprint,dai2023wifi,kordi2024survey,li2022massivemimo,euchner2022bigcsidata}, while acoustic datasets support room acoustics and ultrasonic localization~\cite{chen2024realacousticfields,Echoes_of_accuracy, dataset_ultrasound_techtile}. Existing datasets combining \gls{rf} and acoustic sensing primarily target application-specific tasks such as human activity recognition or drone detection rather than indoor positioning~\cite{mohtadifar2023har,frid2022drone}. Consequently, we are not aware of any publicly available dataset that provides synchronized \gls{rf} and acoustic measurements acquired at identical indoor positions together with accurate ground-truth positioning.

Techtile was designed as a room-scale, distributed, synchronized infrastructure for reproducible experimentation in communication, sensing, positioning, and acoustic research~\cite{techtilePrimer,techtileOpen6g,techtile-acoustic}. Previous Techtile acoustic work and the released ultrasound dataset provide a reusable basis for ultrasonic positioning in the same environment~\cite{Echoes_of_accuracy, dataset_ultrasound_techtile}. ARFT extends these earlier efforts by co-locating an ultrasonic source and an \gls{rf} user antenna on the same rover, producing synchronized \gls{rf}, acoustic and ground-truth position measurements for every acquisition cycle. The resulting release provides common experiment and cycle identifiers across all modalities, eliminating the need for geometric matching between independent measurement campaigns.

The distinguishing feature of ARFT is the synchronized acquisition of distributed \gls{rf} and acoustic measurements in a phase-coherent testbed. Compared with DeepSense~6G~\cite{alkhateeb2023deepsense}, ARFT specifically targets room-scale indoor positioning. Compared with previous Techtile ultrasound data~\cite{Echoes_of_accuracy, dataset_ultrasound_techtile}, it adds co-located \gls{rf} measurements and cycle-level alignment between \gls{rf}, acoustic and position records. Unlike existing \gls{rf}-acoustic datasets developed for activity recognition or detection~\cite{mohtadifar2023har,frid2022drone}, ARFT is designed for localization and enables direct benchmarking of \gls{rf}-only, acoustic-only and joint positioning algorithms. The receiver-side \gls{rf} infrastructure provides relative phase and amplitude across 42 ceiling antennas, supporting spatial fingerprinting and geometry-aware inference, while the acoustic modality preserves raw ultrasonic chirp responses from 91 synchronized microphones for ranging, fingerprinting and multipath analysis. The common synchronization, dense spatial sampling, open parsing scripts and notebook-based processing pipeline additionally support multimodal sensor fusion and cross-modal learning.

ARFT packages a synchronized measurement campaign. It includes the acquisition stack, rover control, acoustic signalling, \gls{rf} measurement scripts, parsing code, processed data and notebook-based analysis workflow used during the campaign. The release follows the broader open-science need for datasets that include data, code and processing material rather than only final benchmark numbers~\cite{overview_datasets_indoor_positioning,dss6g}.

The remainder of the paper is as follows. \Cref{sec:platform} describes the platform and acquisition workflow. \Cref{sec:campaign} quantifies the campaign and its present coverage. \Cref{sec:capabilities} discusses the recorded data and the positioning capabilities as exposed in the analysis workflow. \Cref{sec:discussion} concludes with the main interpretation caveats and current limitations.

\section{Platform and Acquisition Workflow}
\label{sec:platform}

\begin{figure*}[t]
    \centering
    % \begin{subfigure}[b]{0.48\textwidth}
    %     \centering
    %     \includegraphics[width=\linewidth]{figures/techtile-room-overview.jpg}
    %     \caption{Wide view of the Techtile measurement room used for the campaign.}
    % \end{subfigure}
    % \hfill
    \begin{subfigure}[t]{0.48\textwidth}
        \centering
        \includegraphics[height=5cm]{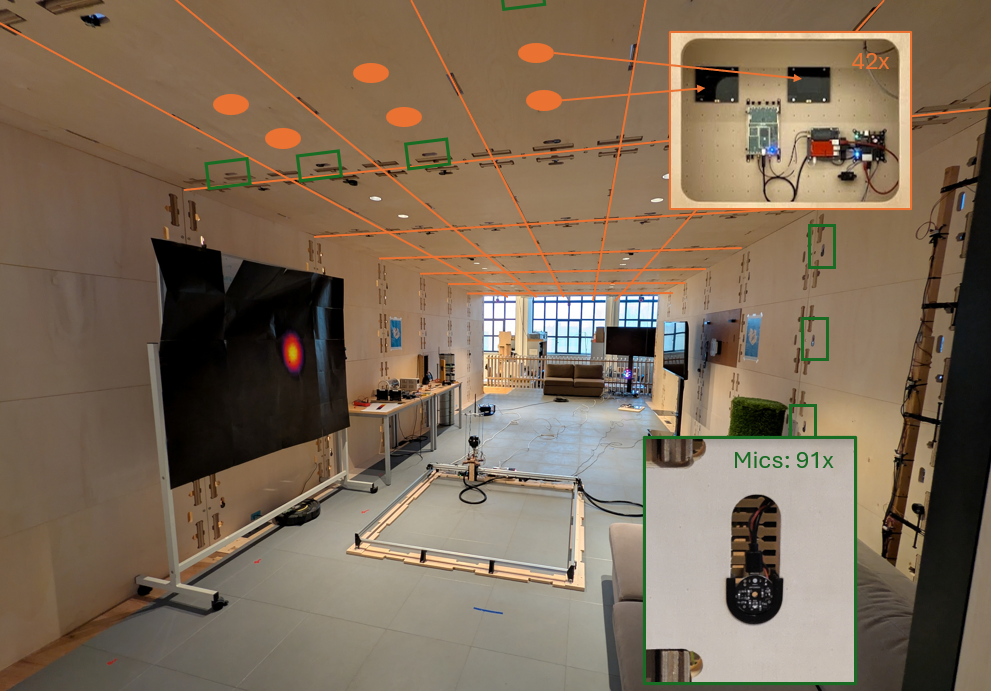}
        \caption{Measurement setup showing the rover frame and room instrumentation.}
    \end{subfigure}\hspace{2cm}%
    \begin{subfigure}[t]{0.32\textwidth}
        \centering
        \includegraphics[height=5cm]{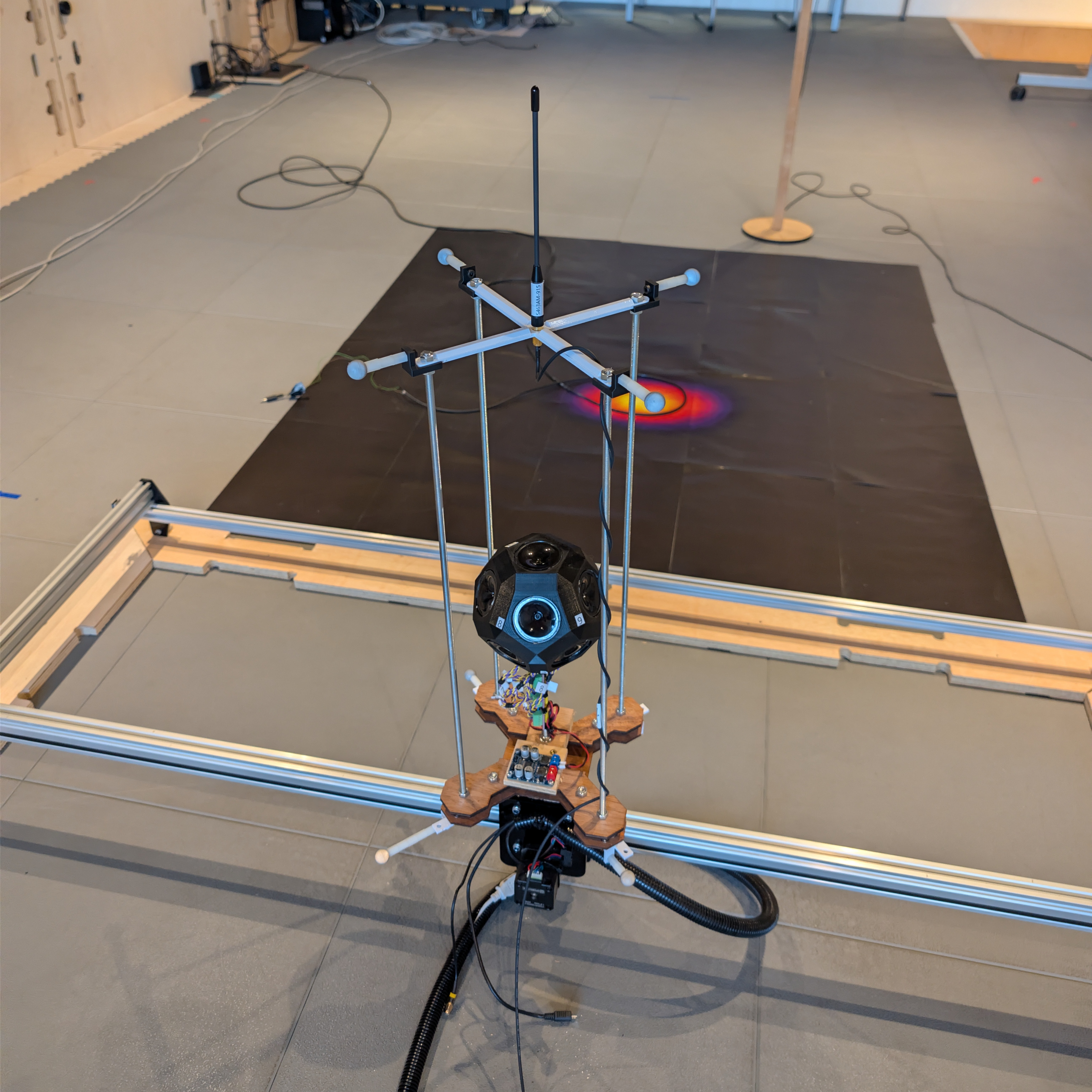}
        \caption{Close-up of the rover-mounted payload combining the \gls{rf} antenna and the omnidirectional acoustic speaker on a common holder.}\label{fig:rover}
    \end{subfigure}
    \caption{Visual overview of the ARFT measurement setup, from room-scale deployment to the co-located RF-acoustic payload on the rover. \vspace{-3mm}}
    \label{fig:measurement-setup-photos}
\end{figure*}

ARFT is collected inside Techtile, a room-scale distributed testbed intended for communication, sensing, positioning, and related 6G experiments~\cite{techtilePrimer,techtileOpen6g, techtile-acoustic}. Techtile is built around 140 detachable tiles with Ethernet-based power, communication, and synchronization. It featurs a Qualisys-based sub-centimeter tracking system, 2D sampling robot, and 42 phase-coherent antennas operating at 920~MHz~\cite{syncWptTechtile,geomWpt2026,dliSuppression2026}. The transmit antenna is not phase-coherent with the receiving antennas.\footnote{This implies that the phase information should be interpreted relative to the other recived phases and should not be considered absolute. For instance, to determine the path length using the phase rotation between the transmitter and receiver.} For the campaign documented in this paper, the measurement rig (\cref{fig:rover}) combines an ultrasonic speaker and an \gls{rf} \gls{ue} antenna on a common mechanical mount so that the acoustic and \gls{rf} channels are sampled at the same physical 2D rover location. Ground-truth position is provided by the Qualisys system. The physical measurement environment is illustrated in \cref{fig:measurement-setup-photos}.

\subsection{Positioning and Measurement Procedure}

The measurements are coordinated by a central orchestrator, that coordinates the movement of the rover, position acquisition and RF and acoustic sampling.

\subsubsection{Orchestator}
The acquisition stack is organized around two control layers. The deployment layer pushes code and settings to the clients and starts the long-lived services. The orchestration layer then coordinates one end-to-end measurement cycle across rover motion, position logging, acoustics, and \gls{rf} synchronization. 
First, the rover moves to the commanded waypoint. Then the system records one fresh ground-truth Qualisys position sample at that stop. Afterward, the acoustic client records one chirp-response capture. Finally, the \gls{rf} orchestrator triggers one synchronized \gls{rf} cycle across the participating radio elements.

\subsubsection{Rover}
\begin{figure}[t]
    \centering
    \includegraphics[width=\linewidth]{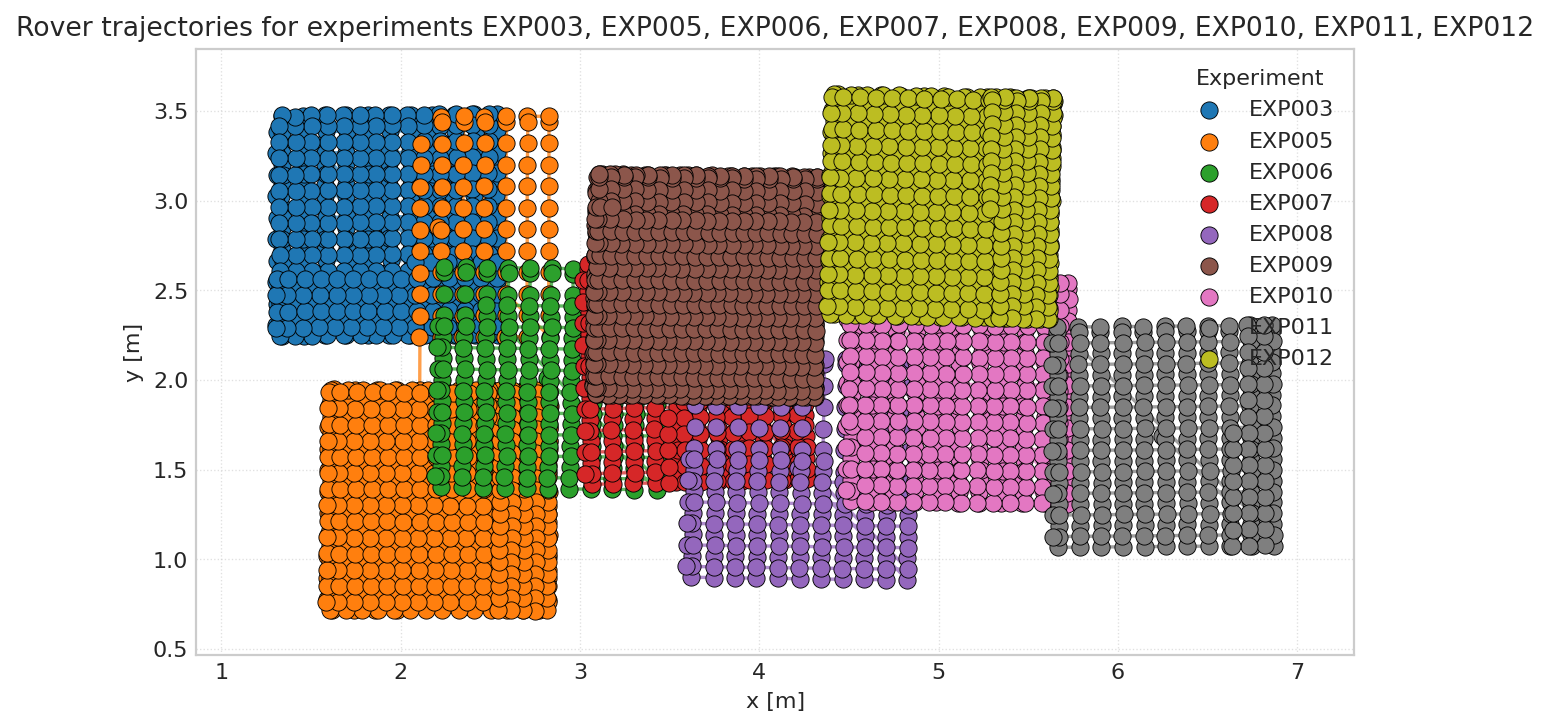}
    \caption{Trajectory and sampled measurement locations. \vspace{-3mm}}\label{fig:samples}
\end{figure}

Both an antenna and speaker are mounted on a rover xy-plotter, as depicted in \cref{fig:rover}. The rover follows a reproducible multi-resolution scan over an $1.23 \times 1.23$~m work area. It performs a sweep over the area and with each iterations the density increases. Hence, the density of the collected samples is related to the total measurement time per area. This can be observed in \cref{fig:samples}. In particular, the rover planner uses a square spiral seven-sweep multi-resolution scan whose spacing is progressively reduced from 120 mm to approximately 50 mm. This makes the measurement process dense enough for spatial visualization while remaining operationally tractable for a multi-modal campaign.

\subsubsection{Positioning Ground Truth}
Ground-truth positioning is provided by the Qualisys system. 6 Miqus M3 cameras are installed, resulting in a 3D standard deviation on the accuracy of 1.25 mm.  After every completed rover move, the orchestrator requests one fresh $(x,y,z)$ position update. Note that in this dataset, the rover only moves in the $(x,y)$ plane, but the $z$-coordinates are stored as well. The reported $z$-coordinate is positioned at the bottom of the transmitter antenna.

\subsection{RF Orchestrator and Measurements}

The active \gls{rf} aperture used by the reciprocity workflow is the Techtile ceiling group: 42 receiver tiles\footnote{
Column \texttt{A} until \texttt{G} and row 5 until 10, covering \texttt{A05} until \texttt{G10}}. The campaign configuration uses the \gls{rf} measurements at 920~MHz with a 250~kS/s sample rate and receiver gain of 80.  

At each rover stop, the \gls{rf} orchestrator waits for an \texttt{ALIVE}  from all the tiles, publishes a  \texttt{SYNC} message, and waits for the corresponding \texttt{DONE} quorum before reporting completion to the main orchestrator. Techtile is further synchronized via a pulse-per-second distribution for time alignment, a shared 10~MHz reference for frequency alignment, and reciprocity-based calibration for front-end phase alignment~\cite{syncWptTechtile}. 

Each receiver stores one pilot-response record per cycle containing the measured pilot phase and amplitude. The reciprocity client estimates a circular-mean pilot phase from the captured IQ samples and pairs it with an \gls{rms} pilot amplitude, post-processing then applies cable-phase correction and converts the result into one complex \gls{csi} value per ceiling antenna. A single physical stop therefore produces a vector of up to 42 phase-coherent complex \gls{csi} observations across the active ceiling receivers. Currently, the dataset is limited to single-band \gls{csi} only.

\subsection{Acoustic Measurements}
The acoustic modality uses the rover-mounted ultrasonic speaker as the active source and the 91 distributed room microphones as fixed receivers. An opposite acoustic dataset in the same environment with the speakers being the anchors and a microphone being the mobile node was used and presented in~\cite{Echoes_of_accuracy, dataset_ultrasound_techtile}. Here, the speaker combines multiple Kemo L010 speakers in a dodecahedron shape forming a quasi-omni-directional ultrasonic speaker. The custom build microphone modules consist of IMP23ABSU \gls{mems} microphones combined with a pure analogue amplifier. One acoustic capture emits a 30~ms linear chirp from 20 to 40~kHz at a 250~kS/s sample rate. The acquisition task records a window twice as long as the excitation so that the direct arrival and the multipath components are preserved in the received signal to make the dataset more usable. The acoustic parameters of the Techtile environment were reported in~\cite{techtile-acoustic}. Techtile is a challenging environment in acoustic terms, with high reverberation and external noise factors. Due to the reverberation time ($\text{RT}_{60}$) of 0.4 s, an acoustic measurement can only take place with this low update rate to exclude interference across measurements. Therefore the acoustic data should be handled as a static dataset. In the current release configuration, these received chirp signals are stored directly rather than being reduced to only deconvolved impulse responses.

On the hardware side, the acoustic hardware shares the same \gls{daq} and thus reference clock, ensuring nanosecond synchronization between output and input tasks. A hardware start triggering ensures that the speaker emission and microphone sampling are aligned within one capture. Each saved acoustic data capture stores the chirp settings together with a microphone label, microphone coordinates, and the recorded waveform samples. The acoustic parser later reorganizes these runtime files into per-experiment waveform tensors indexed by experiment, cycle, microphone label, and sample index. Because the acoustic source and the \gls{rf} antenna are co-located on the rover, the acoustic traces and \gls{rf} snapshots refer to the same physical rover pose, except for a $z$-dimension offset as depicted in Fig.~\ref{fig:measurement-setup-photos} 

\begin{table}[t]
    \centering
    \caption{Key acquisition settings reconstructed from the campaign configuration.}
    \label{tab:config}
    \setlength{\tabcolsep}{4pt}
    \begin{tabular}{p{0.42\columnwidth}p{0.5\columnwidth}}
        \toprule
        Parameter & Value \\
        \midrule
        Testbed & Techtile distributed room-scale infrastructure \\
        Active \gls{rf} aperture & 42 ceiling receiver tiles (\texttt{A05}--\texttt{G10}) \\
        \gls{rf} excitation & Continious waveform pilot \\
        Center frequency & 920 MHz \\
        RF sample rate & 250 kS/s \\
        Receiver gain & 80 \\
        Positioning system & Qualisys \\
        Acoustic rig & ultrasonic speaker co-mounted with the \gls{rf} antenna, 91 microphones in walls and ceiling \\
        Acoustic excitation & 20--40 kHz linear chirp, 30 ms at 250 kS/s \\
        Runtime recording & rover CSV logs with pose, per-host \gls{rf} pilot logs, per-cycle acoustic waveform captures \\
        Rover sweep plan & 7 sweeps with nominal spacings 120, 120, 90, 90, 67.5, 67.5, and 50.6 mm\\
        \gls{rf} sync group & 43 synchronized subscribers in the pilot-synchronization loop \\
        Cycle semantics & 1 rover stop + 1 position sample + 1 acoustic capture + 1 \gls{rf} cycle \\
        \bottomrule
    \end{tabular}
\end{table}

\Cref{tab:config} highlights the key acquisition setting that were used to collect the ARFT dataset. Post-processing begins after acquisition. The \gls{rf} parser scans the per-host result files, extracts complex pilot measurements, applies cable correction, joins the result with rover positions, filters consecutive duplicate positions using a tolerance derived from the rover configuration, and writes one merged NetCDF. The acoustic parser produces a separate per-experiment NetCDF of recorded waveforms. ARFT therefore preserves both the runtime logs and the processed analysis data. The latter ultimately forms the merged dataset.

\section{Experimental Campaign and Coverage}
\label{sec:campaign}

\begin{table}[t]
    \centering
    \caption{Per-experiment campaign summary detailing the number of collected samples per type.}
    \label{tab:experiments}

    \scriptsize
    \setlength{\tabcolsep}{3pt}

    \begin{tabular}{lcccccc}
        \toprule
        Experiment &
        $x$ range &
        $y$ range &
        \begin{tabular}{@{}c@{}}Rover\\cycles\end{tabular} &
        \begin{tabular}{@{}c@{}}\gls{rf}\\cycles\end{tabular} &
        \begin{tabular}{@{}c@{}}Acoustic\\cycles\end{tabular} &
        \begin{tabular}{@{}c@{}}Tri-modal\\cycles\end{tabular} \\
        \midrule

        EXP003 & [1.31, 2.54] & [2.25, 3.48] & 529  & 529  & 529  & 529 \\
        EXP005 & [1.58, 2.83] & [0.71, 3.47] & 777  & 777  & 1618 & 777 \\
        EXP006 & [2.19, 3.44] & [1.39, 2.63] & 238  & 238  & 238  & 238 \\
        EXP007 & [3.01, 4.27] & [1.42, 2.68] & 378  & 378  & 379  & 378 \\
        EXP008 & [3.59, 4.84] & [0.89, 2.13] & 201  & 201  & 201  & 201 \\
        EXP009 & [3.07, 4.32] & [1.90, 3.15] & 1355 & 1355 & 1356 & 1355 \\
        EXP010 & [4.48, 5.73] & [1.31, 2.55] & 458  & 458  & 458  & 458 \\
        EXP011 & [5.63, 6.87] & [1.07, 2.31] & 291  & 291  & 291  & 291 \\
        EXP012 & [4.38, 5.64] & [2.34, 3.60] & 784  & 784  & 785  & 784 \\

        \bottomrule
    \end{tabular}
\end{table}

The ARFT dataset campaign contains 12 named experiments (\texttt{EXP001}--\texttt{EXP012}). Across those runs, the recorded rover coordinates span approximately 0.81~m to 6.87~m in the Qualisys $x$-axis and 0.71~m to 3.60~m in the $y$-axis. The experiments are not all mapped to the same spatial subregion, which is evident from the per-experiment coordinate ranges in \cref{tab:experiments} and \cref{fig:samples}. The ARFT dataset contains 5011 tri-modal spatial samples. As can be observed, not all experiment runs, constitute the same number of samples and density.
% This corrected tri-modal subset spans 5.57~m in the $x$ direction and 2.89~m in the $y$ direction, corresponding to a 16.09~m$^2$ axis-aligned bounding box.

% Several observations follow directly from \cref{tab:experiments}. First, the campaign is not uniformly complete. EXP003, EXP006, EXP008, EXP010, and EXP011 are fully aligned. EXP007, EXP009 and EXP012 are nearly complete, each missing only one \gls{rf} and rover cycle. EXP005 contributes a large acoustic subset, but lacks \gls{rf} and rover cycles. Finally, EXP001, EXP002, and EXP004 provide rover coverage but not processed \gls{rf} or acoustic cycles in the present release.

% The campaign should therefore be interpreted as a layered asset rather than a single homogeneous benchmark. If the research question is purely acoustic-spatial, the usable set is materially larger than if joint \gls{rf}+acoustic analysis is required.

\section{Recorded Data Products and Positioning Capabilities}
\label{sec:capabilities}

ARFT exposes several distinct data products, each corresponding to a different layer of the workflow. At acquisition time, the rover logger writes one CSV row per completed move, the \gls{rf} clients write one runtime record per hostname and cycle, and the \gls{rf} orchestrator writes one YAML summary entry per synchronized cycle. The post-processing path then converts these operational files into analysis-ready NetCDF products. The use of hierarchical NetCDF products for the released tensors is also consistent with broader 6G testbed efforts that standardize reusable measurement datasets on top of NetCDF/HDF5 storage backends~\cite{dss6g}.

The current merged \gls{rf} NetCDF aggregates EXP003 and EXP005--EXP012 and has dimensions \texttt{experiment\_id} $\times$ \texttt{cycle\_id} $\times$ \texttt{hostname}. Its main variables are \texttt{csi\_real}, \texttt{csi\_imag}, \texttt{csi\_available}, \texttt{rover\_x}, \texttt{rover\_y}, \texttt{rover\_z}, and \texttt{position\_available}. One physical \gls{rf} observation is reconstructed as \texttt{csi\_real} $+ j\,\texttt{csi\_imag}$, while \texttt{csi\_available} indicates whether a given \texttt{(experiment\_id, cycle\_id, hostname)} tuple is populated. This design gives the user both direct array access and an explicit validity mask.

The \gls{rf} product should be interpreted as a distributed spatial fingerprint of the rover location. Each cycle yields a 42-element complex vector over the ceiling tiles, derived from the per-host pilot phase and amplitude measurements after cable correction. Across the 5011 populated \gls{rf} cycles in the merged file, the dataset stores 210068 tile-level complex \gls{csi} samples. Host coverage is also strong: 4617 of the 5011 \gls{rf} cycles contain all 42 ceiling-tile entries, and the remaining 394 cycles contain 41 entries. In other words, the typical measurement point is not a sparse partial vector but a nearly complete 42-element spatial snapshot. This makes the \gls{rf} modality directly usable for fingerprint-based positioning, for geometry-aware inference that exploits the distributed aperture, and for learning spatial channel maps that are later inverted for localization.

The acoustic product complements this with waveform-level information. The acoustic pipeline writes one NetCDF per experiment with waveform data organized on \texttt{(experiment\_id, cycle\_id, microphone\_label, sample\_index)}. Each stored trace corresponds to the chirp response recorded by a microphone at the same rover stop as the paired \gls{rf} snapshot. The acoustic modality therefore preserves time-domain propagation structure such as direct-path timing, reverberation, and multipath signatures. These traces can be used for acoustic fingerprinting, feature extraction for position regression or classification, or model-based processing that uses arrival-time and room-response cues for localization.

\begin{figure}[t]
    \centering
    \includegraphics[width=0.5\textwidth]{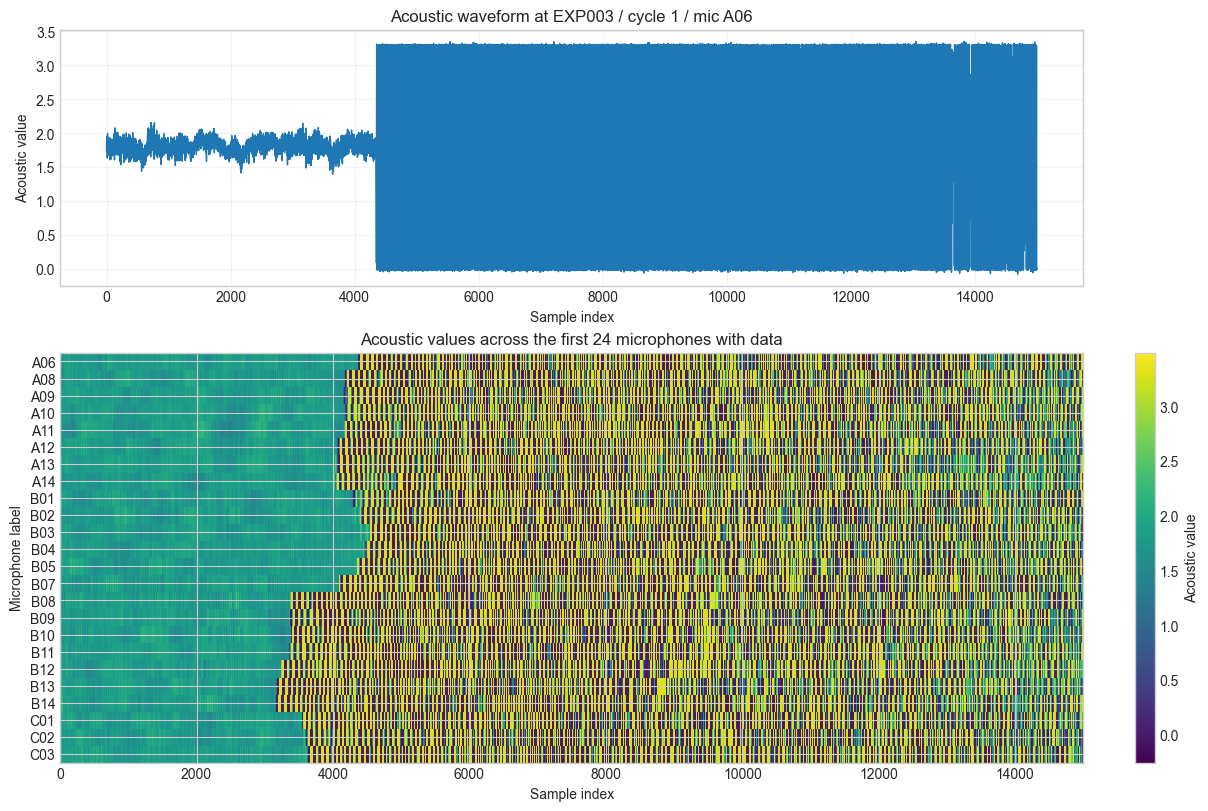}
    \caption{Acoustic waveform inspection for the position resolved by the joint \gls{rf}-acoustic notebook. The upper panel shows one microphone trace for \texttt{EXP003}/cycle~1, while the lower panel shows the corresponding waveform matrix for the first microphones with available data at the same acquisition cycle. \vspace{-3mm}}
    \label{fig:rf-acoustic-position-waveforms}
\end{figure}

\begin{figure}[t]
    \centering
    \begin{subfigure}[t]{0.48\textwidth}
        \centering
        \includegraphics[width=\linewidth]{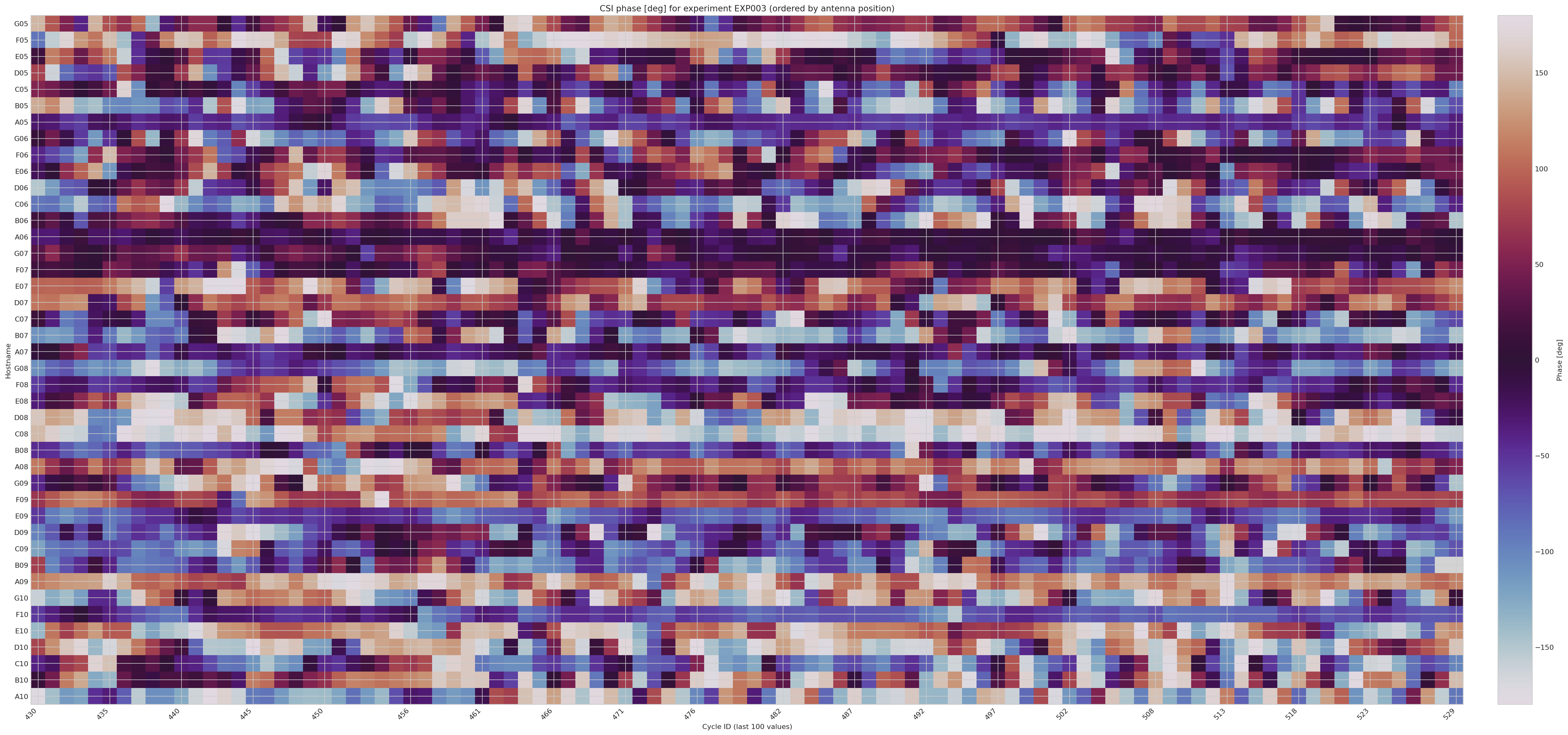}
        \caption{\gls{csi} phase over the selected cycle window.}
    \end{subfigure}\hfill%
    \begin{subfigure}[t]{0.48\textwidth}
        \centering
        \includegraphics[width=\linewidth]{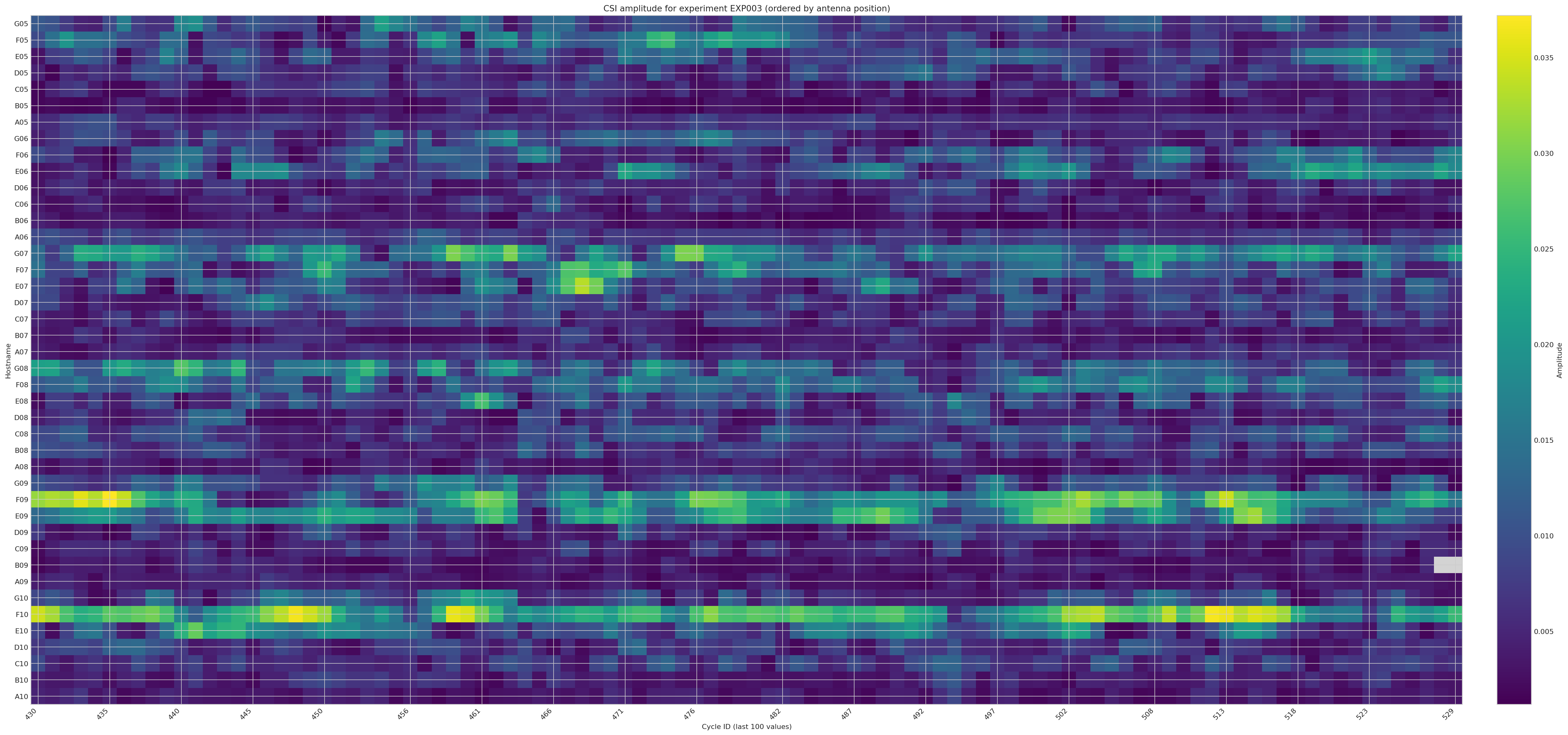}
        \caption{\gls{csi} amplitude over the same cycle window.}
    \end{subfigure}
    \caption{Heatmaps from \texttt{tutorial\_csi\_per\_position.ipynb}. The rows are ordered ceiling receiver hostnames, and the columns show the final cycle IDs in the selected experiment window. Together they expose stable receiver-dependent structure and temporal variation in the \gls{rf} channel tensor. \vspace{-3mm}}
    \label{fig:csi-position-heatmaps}
\end{figure}

The key capability is tri-modal alignment. Because the acoustic and \gls{rf} parsers use the same \texttt{(experiment\_id, cycle\_id)} key, ARFT supports direct cycle-level joining between rover position, tile-level \gls{rf} channel measurements, and raw microphone waveforms. Since each cycle also has a ground-truth rover pose from Qualisys, the release supports three natural positioning regimes: RF-only positioning, acoustic-only positioning, and joint multimodal positioning in which both feature sets are fused against the same $(x,y,z)$ target. \Cref{fig:rf-acoustic-position-waveforms,fig:csi-position-heatmaps} show two notebook-derived views of this alignment and \gls{rf} structure: one cycle-level acoustic waveform matrix joined to a selected \gls{rf} position, and phase/amplitude heatmaps over a sequence of \gls{rf} cycles.

\begin{figure}[]
    \centering
    \includegraphics[width=\linewidth]{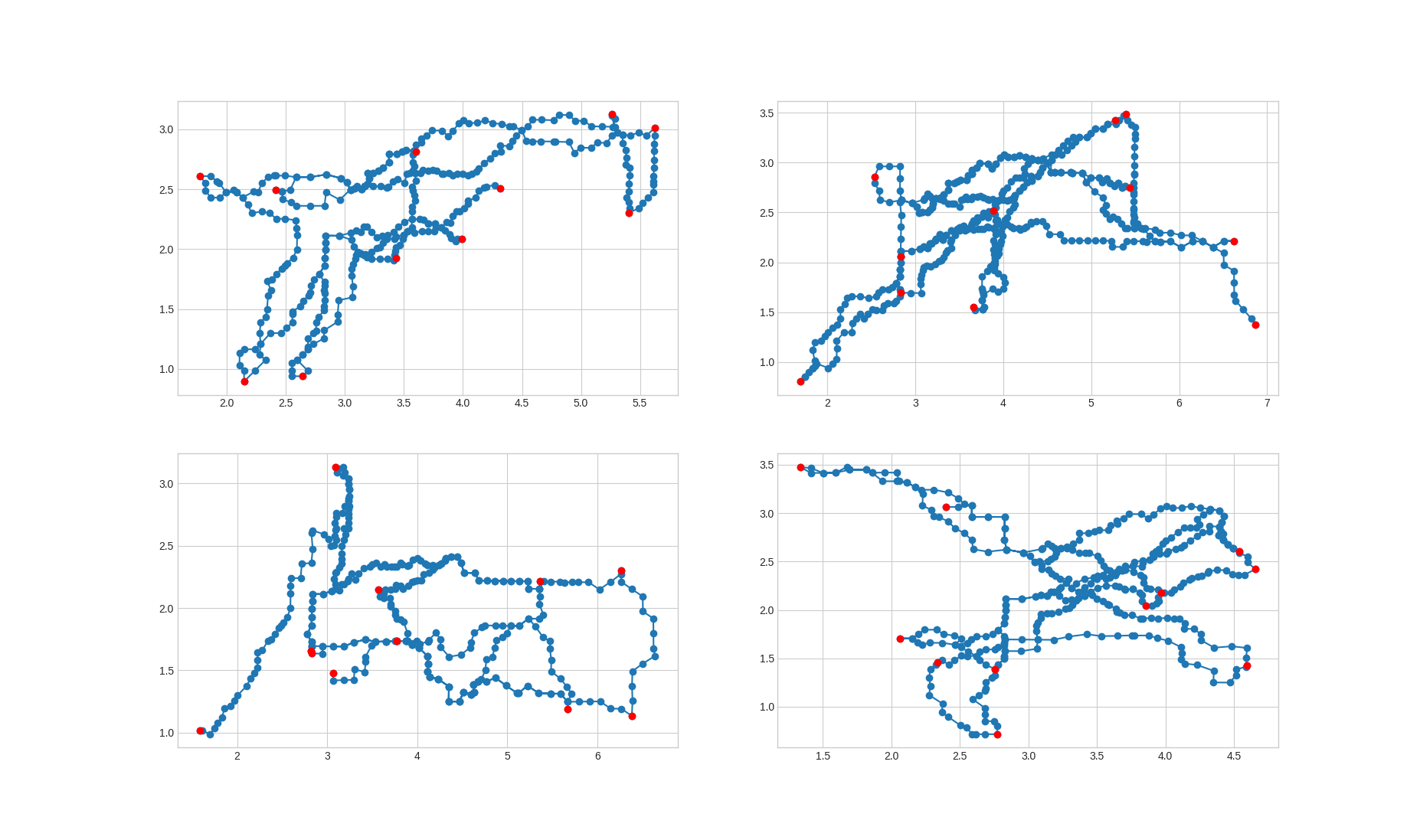}
    \caption{Four randomly generated walks. Red indicates a static point, and blue indicates points on the paths between static points. \vspace{-3mm}}
    \label{fig:generatedwalks}
\end{figure}

\begin{figure}[]
    \centering
\includegraphics[width=0.4\textwidth]{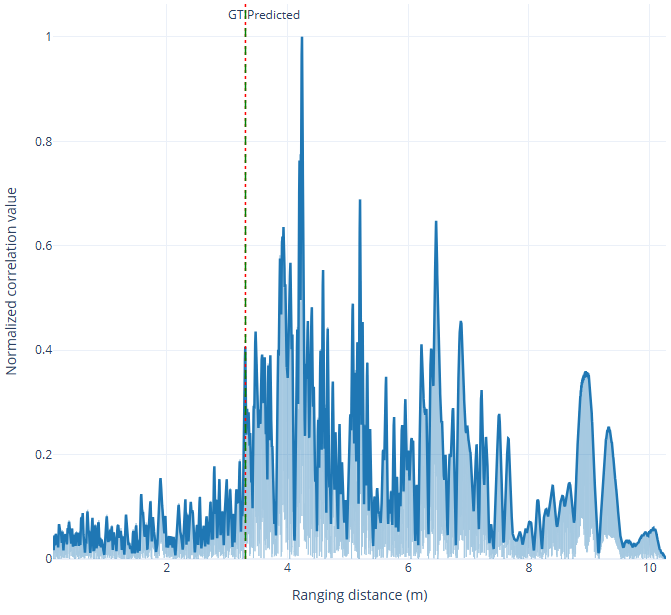}
    \caption{Ranging estimate example for cycle 304 EXP007 microphone C04 with the correlation values and LPF values. The correct peak resulting in the estimated range is determined using the peak-prominence algorithm. The estimated and ground-truth range is depicted. \vspace{-3mm}}
    \label{fig:correlation_graph_example}
\end{figure}

The post-processing workspace now makes the RF--acoustic positioning path concrete rather than merely hypothetical. It includes a graph-based walk generator and persisted train/test trajectory splits, with 2500 generated training walks and 50 held-out test walks. The test set positions are excluded from the training walks data.
The walk generator works by choosing a sequence of static flagged positions, and repeatedly performing Dijkstra's algorithm to find the shortest path from one static point to the next, along the rover measurement points. Static positions are introduced to perform acoustic positioning. The sequence of static points can either be randomly generated or manually specified. There is also an option to forbid the paths from containing specific points. Figure \ref{fig:generatedwalks} showcases four examples of randomly generated walks.

The acoustic positioning scripts operate on the static flagged positions in the paths by filtering the received signals, selecting anchor microphones, applying chirp pulse compression as depicted in Fig.~\ref{fig:correlation_graph_example}, estimating time-of-arrival ranges based on a peak-prominence algorithm, and by determining a least-squares position estimate for each rover stop. The most optimal peak-prominence-factor is determined for this dataset, resulting in a value equal to 0.24.  The current configuration uses up to 50 selected anchors per stop with sum-rate-based ranking due to the saturating operation mode of the microphones, which means that the released waveform data is already exercised by a full end-to-end model-based localization pipeline. 

\begin{figure}[t]
    \centering
    \includegraphics[width=\linewidth]{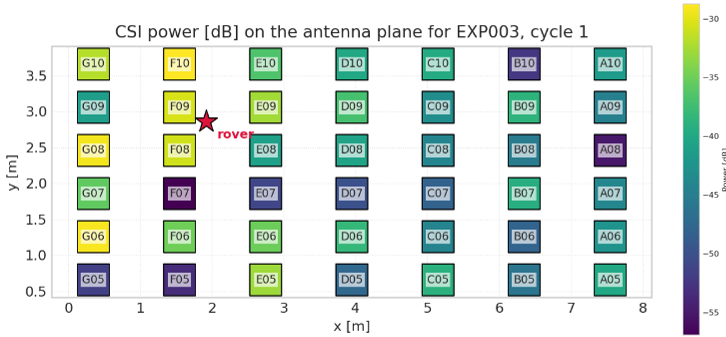}
    \caption{Example per-position \gls{rf} power snapshot over the 42 ceiling receivers. }\label{fig:capability-artifacts}
\end{figure}

The saved evaluation artifacts are strong enough to summarize as experimental evidence. The calibrated acoustic results cover 50 held-out paths and 550 localized rover stops. Across that set, the mean 2D positioning error is 0.110~m and the P90 value is equal to 0.195~m, while the mean 3D positioning error is 0.130~m and the P90 value is equal to 0.217~m. The accompanying ranging records contain 27500 anchor-wise distance estimates with mean absolute error 0.311~m. These values should be interpreted as documented baseline performance of the current model-based acoustic pipeline, not as an upper bound on what the dataset can support. At present, the \gls{rf} pipeline regarding positioning is still exploratory compared with the more complete acoustic ranging-and-localization path. The lack of phase synchronization between the RF transmitter and receivers makes accurate phase-based RF positioning particularly challenging. \Cref{fig:capability-artifacts} illustrates tile-level interpretation of one synchronized \gls{rf} capture. This is a meaningful capability distinction for users who may be interested in fingerprinting, distributed channel modeling, RF-acoustic fusion, or direct positioning inference.

The analysis workflow already demonstrates the most immediate downstream uses. The provided notebooks support:
\begin{itemize}
    \item campaign-level trajectory inspection and measurement-cloud visualization
    \item overview heatmaps of \gls{csi} phase and amplitude over the measurement set
    \item single-position spatial snapshots across the 42 ceiling receivers
    \item joint \gls{rf}-acoustic inspection at a selected rover stop
    \item export of movie sequences showing spatial evolution of phase and power.
\end{itemize}
Taken together with the acoustic and \gls{rf} post-processing scripts, these are not promised future tasks, they are concrete analyses already represented by scripts, notebooks, and exported figures in the present project release.

\section{Discussion and Interpretation Caveats}
\label{sec:discussion}

The ARFT dataset's main strength is not just its multi-modal measurements, but its explicit distinction between the logged campaign and the analysis-ready subset. As a result, transparency about the currently merged 9 (rather than 12) experiments is essential. The completeness report should therefore be read on two levels: rover logs show what was physically traversed, while the merged \gls{rf}-acoustic dataset shows what is available for channel analysis. This is why EXP001, EXP002, and EXP004 remain relevant to the campaign narrative, despite not contributing processed \gls{rf} and acoustic data to the current merged set.

The second caveat is structural. The \texttt{cycle\_id} axis in the merged \gls{rf} NetCDF is a shared union axis across experiments, not a guarantee that every experiment occupies every cycle index. Validity is encoded by the availability masks, not merely by coordinate presence. Any downstream study that ignores \texttt{csi\_available} or \texttt{position\_available} risks accidentally treating absent data as zero-valued data or assuming that coordinate existence implies measurement existence.

Furthermore, the measurements were performed statically. Therefore, especially the slow-propagating and long reverberating acoustic waves should be handled as static measurements to perform realistic positioning. Besides, due to the hardware availability this campaign could only be performed in the Techtile environment. Scaling to other rooms requires hardware re-design.

\section{Conclusion}

This paper presents ARFT as a synchronized measurement campaign and data release rather than a benchmark. It integrates rover motion with Qualisys positioning, acoustic sensing, and distributed ceiling-tile \gls{rf} measurements in Techtile. The current release comprises 12 experiments (6735 rover stops), including a merged 9-experiment \gls{rf} dataset (5011 cycles, 210068 \gls{csi} values), 5855 acoustic cycles, and a fully aligned tri-modal subset of 5011 cycle pairs.

The key contribution is clarity: ARFT supports spatial analysis, tile-level \gls{rf} inspection, multimodal cycle alignment, and baseline acoustic localization, provided the appropriate completeness subset is used. By explicitly documenting coverage, processing, example scripts and limitations, this work establishes a reliable dataset for distributed communication and multimodal sensing research~\faicon{github}\footnote{\url{https://github.com/techtile-by-dramco/ELLIIIT-dataset-26/}}.

\section*{Acknowledgment}
This work was supported by the AMBIENT-6G project, which received funding from the Smart Networks and Services Joint Undertaking (SNS JU) under the European Union’s Horizon Europe research and innovation programme under Grant Agreement No. 101192113. This work was supported by the IoF fund of KU Leuven allocated to the `POMPOEN' project. Furthermore, a special thanks to the ELLIIT community to support us and to provide the inputs to create this dataset. Thank you to Geoffrey Ottoy and Bert Cox for the collaboration with the realization of the Techtile infrastructure.

\printbibliography%

@misc{dliSuppression2026,
	title        = {{Experimental Study of Interference Suppression for Backscatter Communication in Distributed MIMO}},
	author       = {Ahmet Kaplan and Gilles Callebaut and Jarne Van Mulders and Erik G. Larsson},
	year         = 2026,
	eprint       = {2604.09061},
	archiveprefix = {arXiv}
}

@inproceedings{dss6g,
	title        = {{An Open Dataset Storage Standard for 6G Testbeds}},
	author       = {Callebaut, Gilles and Sandra, Michiel and Nelson, Christian and Wilding, Thomas and Delabie, Daan and Deutschmann, Benjamin J. B. and Tärneberg, William and Fitzgerald, Emma and Johansson, Anders J. and Van der Perre, Liesbet},
	year         = 2023,
	booktitle    = {2023 IEEE Conference on Antenna Measurements and Applications (CAMA)},
	volume       = {},
	number       = {},
	pages        = {347--352}
}

@misc{geomWpt2026,
	title        = {{Experimental Evaluation of Geometry and Reciprocity-Based Beamforming with Large Arrays for RF Wireless Power Transfer}},
	author       = {Gilles Callebaut and Jarne Van Mulders},
	year         = 2026,
	booktitle    = {2026 IEEE Wireless Power Technology Conference and Expo (WPTCE)}
}

@inproceedings{syncWptTechtile,
	title        = {{Experimental Study on the Effect of Synchronization Accuracy for Near-Field RF Wireless Power Transfer in Multi-Antenna Systems}},
	author       = {Callebaut, Gilles and Van Mulders, Jarne and Cox, Bert and Deutschmann, Benjamin J. B. and Ottoy, Geoffrey and De Strycker, Lieven and Van der Perre, Liesbet},
	year         = 2025,
	booktitle    = {2025 19th European Conference on Antennas and Propagation (EuCAP)},
	volume       = {},
	number       = {},
	pages        = {01--05}
}

@inproceedings{techtileOpen6g,
	title        = {{Techtile -- Open 6G {R\&D}  Testbed for Communication, Positioning, Sensing, WPT and Federated Learning}},
	author       = {Callebaut, Gilles and Mulders, Jarne Van and Ottoy, Geoffrey and Delabie, Daan and Cox, Bert and Stevens, Nobby and Perre, Liesbet Van der},
	year         = 2022,
	booktitle    = {2022 Joint European Conference on Networks and Communications {\&} 6G Summit (EuCNC/6G Summit)},
	volume       = {},
	number       = {},
	pages        = {417--422}
}

@article{techtilePrimer,
	title        = {{A Primer on Techtile: An {R\&D} Testbed for Distributed Communication, Sensing and Positioning}},
	author       = {Gilles Callebaut and Jarne Van Mulders and Geoffrey Ottoy and Liesbet Van der Perre},
	year         = 2021,
	journal      = {arXiv preprint arXiv:2105.06740},
}

@inproceedings{techtile-acoustic,
	title        = {{Techtile: a Flexible Testbed for Distributed Acoustic Indoor Positioning and Sensing}},
	author       = {Delabie, Daan and Cox, Bert and De Strycker, Lieven and Van der Perre, Liesbet},
	year         = 2022,
	booktitle    = {IEEE Sensors Applications Symposium (SAS)}
}

@article{Echoes_of_accuracy,
	title        = {{Echoes of Accuracy: Enhancing Ultrasonic Indoor Positioning for Energy-Neutral Devices With Neural Network Approaches}},
	author       = {Delabie, Daan and Feys, Thomas and Buyle, Chesney and Cox, Bert and Van der Perre, Liesbet and De Strycker, Lieven},
	year         = 2025,
	journal      = {IEEE Journal of Indoor and Seamless Positioning and Navigation},
	volume       = 3,
	number       = {},
	pages        = {227--244}
}

@article{overview_datasets_indoor_positioning,
	title        = {{Comprehensive Assessment of Open Science Practices in Indoor Positioning: Open Data, Code, and Material}},
	author       = {Anagnostopoulos, Grigorios G. and Barsocchi, Paolo and Crivello, Antonino and Pendão, Cristiano and Silva, Ivo and Torres-Sospedra, Joaquín},
	year         = 2025,
	journal      = {IEEE Journal of Indoor and Seamless Positioning and Navigation},
	volume       = 3,
	number       = {},
	pages        = {175--194}
}

@dataset{dataset_ultrasound_techtile,
	title        = {{Ultrasound Indoor Positioning Dataset}},
	author       = {Delabie, Daan},
	year         = 2025,
	month        = feb,
	publisher    = {Zenodo},
	doi          = {10.5281/zenodo.14833225},
	url          = {https://doi.org/10.5281/zenodo.14833225},
	version      = {0.0.1.}
}

@article{alkhateeb2023deepsense,
	title        = {{DeepSense 6G: A Large-Scale Real-World Multi-Modal Sensing and Communication Dataset}},
	author       = {Alkhateeb, Ahmed and others},
	year         = 2023,
	journal      = {IEEE Communications Magazine},
	volume       = 61,
	number       = 9,
	pages        = {122--128}
}

@article{li2022massivemimo,
	title        = {{Toward Fine-Grained Indoor Localization Based on Massive MIMO-OFDM System: Experiment and Analysis}},
	author       = {Li, Chenglong and De Bast, Sibren and Tanghe, Emmeric and Pollin, Sofie and Joseph, Wout},
	year         = 2022,
	journal      = {IEEE Sensors Journal},
	volume       = 22,
	number       = 6,
	pages        = {5318--5328}
}

@inproceedings{euchner2022bigcsidata,
	title        = {{A Distributed Massive MIMO Channel Sounder for Big CSI Data-driven Machine Learning}},
	author       = {Euchner, Florian and Gauger, Marc and Doerner, Sebastian and ten Brink, Stephan},
	year         = 2021,
	booktitle    = {WSA 2021; 25th International ITG Workshop on Smart Antennas},
	volume       = {},
	number       = {},
	pages        = {1--6}
}

@inproceedings{kazemi2022csifingerprint,
	title        = {{User-Side Indoor Localization Using CSI Fingerprinting}},
	author       = {Kazemi, Parham and Al-Tous, Hanan and Studer, Christoph and Tirkkonen, Olav},
	year         = 2022,
	booktitle    = {2022 IEEE 23rd International Workshop on Signal Processing Advances in Wireless Communication (SPAWC)},
	volume       = {},
	number       = {},
	pages        = {1--5}
}

@article{kordi2024survey,
	title        = {{Survey of Indoor Localization Based on Deep Learning}},
	author       = {Khaldon Azzam Kordi and Mardeni Roslee and Mohamad Yusoff Alias and Abdulraqeb Alhammadi and Athar Waseem and Anwar Faizd Osman},
	year         = 2024,
	journal      = {Computers, Materials and Continua},
	volume       = 79,
	number       = 2,
	pages        = {3261--3298}
}

@article{dai2023wifi,
	title        = {{A Survey of Latest Wi-Fi Assisted Indoor Positioning on Different Principles}},
	author       = {Dai, Jihan and Wang, Maoyi and Wu, Bochun and Shen, Jiajie and Wang, Xin},
	year         = 2023,
	journal      = {Sensors},
	volume       = 23,
	number       = 18,
	article-number = 7961
}

@INPROCEEDINGS{chen2024realacousticfields,
  author={Chen, Ziyang and Gebru, Israel D. and Richardt, Christian and Kumar, Anurag and Laney, William and Owens, Andrew and Richard, Alexander},
  booktitle={2024 IEEE/CVF Conference on Computer Vision and Pattern Recognition (CVPR)}, 
  title={{Real Acoustic Fields: An Audio-Visual Room Acoustics Dataset and Benchmark}}, 
  year={2024},
  volume={},
  number={},
  pages={21886-21896},
}

@Article{mohtadifar2023har,
AUTHOR = {Mohtadifar, Masoud and Cheffena, Michael and Pourafzal, Alireza},
TITLE = {{Acoustic- and Radio-Frequency-Based Human Activity Recognition}},
JOURNAL = {Sensors},
VOLUME = {22},
YEAR = {2022},
NUMBER = {9},
ARTICLE-NUMBER = {3125},
}

@Article{frid2022drone,
AUTHOR = {Frid, Alan and Ben-Shimol, Yehuda and Manor, Erez and Greenberg, Shlomo},
TITLE = {{Drones Detection Using a Fusion of RF and Acoustic Features and Deep Neural Networks}},
JOURNAL = {Sensors},
VOLUME = {24},
YEAR = {2024},
NUMBER = {8},
ARTICLE-NUMBER = {2427},
}

\end{document}